\documentclass[twocolumn,prl,showpacs,amsmath,amssymb,floatfix,superscriptaddress]{revtex4-1}
\usepackage[latin9]{inputenc}
\usepackage{color}
\usepackage{array}
\usepackage{multirow}
\usepackage{graphicx}
\usepackage{wasysym}
\usepackage{esint}

\usepackage{xcolor}
\definecolor{zgreen}{rgb}{0.0, 0.26, 0.15}

\makeatletter

\providecommand{\tabularnewline}{\\}

\pdfoutput=1

\usepackage{dcolumn}
\usepackage{bm}

\@ifundefined{showcaptionsetup}{}{%
 \PassOptionsToPackage{caption=false}{subfig}}
\usepackage{subfig}
\makeatother

\begin{document}
\title{A hybrid approach to basis set independent Poisson solver for an arbitrary
charge distribution}
\author{Po-Hao Chang}
\email{pchang2@utep.edu}
\affiliation{Physics Department, University of Texas at El Paso, El Paso, Texas 79968, USA}

\author{Zachary Buschmann}
\affiliation{Physics Department, University of Texas at El Paso, El Paso, Texas 79968, USA}

\author{Rajendra R. Zope}
\email{rzope@utep.edu}
\affiliation{Physics Department, University of Texas at El Paso, El Paso, Texas
79968, USA}

\begin{abstract}
We review two common numerical schemes for Coulomb potential evaluation
that differ only in their radial part of the solutions in the spherical
harmonic expansion (SHE). One is based on finite-difference method
(FDM) while the other is based on the Green's function (GF) solution
to the radial part of the Poisson equation. We analyze the methods
and observe that the FDM-based approach appears to be more efficient
in terms of the convergence with the number of radial points, particularly
for monopole ($l=0)$. However, as a known issue, it suffers from 
error accumulation as the system size increases. We identify the source
of error that comes mainly from $l=1$ (and sometimes $l=2$)
contribution of SHE induced by the charge partitioning. We then propose
a hybrid scheme by combining the two methods, where the radial solution 
for $l=0$ is obtained using the FDM method and 
treating the remaining terms using GF approach.
The proposed hybrid method is subsequently applied to a variety of
systems to examine its performance. The results show improved accuracy
than earlier numerical schemes in all cases. 
We also show that, even
with a generic set of radial grid parameters, accurate energy differences
can be obtained using a numerical Coulomb solver in standard density
functional studies.
\end{abstract}
\maketitle

\section{Introduction}

The rapid advancement of computing power has allowed accurate
numerical simulations of material properties for more realistic and
complex systems under first-principles-based approaches. Although
DFT is the most widely applied quantum mechanical method, as it 
offers the best balance
between the accuracy and the computational cost, the evaluation of
Hartree contribution to the Hamiltonian that describes the classical part of electron-electron
interactions remains one of the major numerical challenges. Therefore, it is
crucial to be able to determine Hartree energy efficiently and accurately. 

In of computational quantum chemistry, Gaussian type orbitals
(GTOs) are among the most popular basis sets as they benefit from
the analyticity of Gaussian product rules and of their integrals which
allows various integrals like the basis overlap integrals, Coulomb
integral, exchange integrals, dipole integrals etc. to be evaluated
efficiently and analytically \cite{helgaker2014molecular}.

Over the years, many efficient schemes have been developed around GTOs
to tackle computationally costly two-electron integral in Coulomb problems.
\cite{mcmurchie1978one,obara1986efficient,friesner1985solution,dunlap1979some,dunlap1983fitting,
mintmire1982fitting,dunlap2000robust,Termath1994b,cp975667rf321987fast,greengard1988rapid,appel1985efficient,barnes1986hierarchical,strain1996achieving,schmidt1991implementing,Becke1988e,Delley1990b}
 Starting from the Boys function for Coulomb potential due
to a spherically symmetric Gaussian charge distribution, the potential
due to charge involving higher angular momentum components can
be efficiently constructed recursively. The most common recursive
schemes are the ones proposed by McMurchie and Davidson \cite{mcmurchie1978one}
and Obara and Saika \cite{obara1986efficient}. These common GTO integral evaluations 
have been implemented in integrals libraries such as libint and libcint \cite{Libint2,libcint2015} 
which are at the heart of many software packages.

Despite the fact that the recursive schemes improve the performance
significantly, the poor scaling of 4-center integrals is still inherently problematic. 
In order to achieve a better scaling, followig up on the idea of combining
numerical grid and a finite basis set \cite{friesner1985solution},
Termath and Handy proposed to analytically evaluate the Coulomb potential
on a numerical grid \cite{Termath1994b}. Not only does this approach
drastically reduce the complexity, the idea can also be further combined
with either standard far-field expansion \cite{Jackson:100964} or
efficient and accurate algorithm such as faster multipole method
FMM and tree codes \cite{cp975667rf321987fast,greengard1988rapid,appel1985efficient,barnes1986hierarchical,strain1996achieving,schmidt1991implementing}.
The advantage becomes particularly obvious when it involves higher angular
momentum basis functions.

A pure numerical basis-set-free approach to the Coulomb problem in
polyatomic systems was first proposed by Becke and Dickson \cite{Becke1988d}.
The method utilizes multi-center numerical grid \cite{Becke1988e},
proposed also by Becke the same year, that partitions the space into
multiple regions. Such partitioning of space allows the integration 
within each region to be performed independently by solving the 
Poisson equations  for the charge density confined in each region. 
By combining with multipolar expansions, one can then obtain 
a set of radial part of the Poisson equations. Once the radial
equations are solved, the solutions can be used to construct 
Coulomb potential in the form of spherical harmonic expansion (SHE).

The approach is appealing but unfortunately suffers from accuracy 
issues when it comes to more spatially extended large molecules 
as the error per atom appears to be accumulative \cite{Termath1994b}. This limits the 
the method from being applied to larger size systems. The extensive 
literature search revealed that 
only a handful of implementations \cite{Posada2018b,Shiozaki2007b,Termath1994b,becke1989basis}
adopted this method and majority of these works  only focused on small molecules
\cite{becke1989basis,becke2016vertical,becke2018communication,becke2019dependence,Shiozaki2007b}. 

Shortly after, based on the same theoretical framework, Delley used
the integral form of the Green's function (GF) solution to the Laplacian
\cite{Delley1990b} as an alternative to solve the radial part of
the Poisson equations. This approach removes the problem of error 
accumulation with system size, thereby opening a gateway 
to broader numerical Poisson applications. The method
is now more widely used and has been implemented in a few DFT codes
\cite{Franchini2014b,Blum2009b} for large scale calculations.

The numerical Poisson solver based on multi-center grid and multipolar
expansion has been discussed in several earlier works and shown promises.
But it is mostly adopted in grid \cite{becke1989basis}, Slater-type
orbital (STO) \cite{te2001chemistry,Franchini2013b} or numerical
atomic orbital (NAO) based codes \cite{Blum2009b,te1991precise}, and
rarely applied to Gaussian-based DFT codes as the Coulomb potential
can already be calculated analytically \cite{obara1986efficient,mcmurchie1978one}.
There have been only a handful of earlier works that attempted at
applying the numerical Poisson solver to GTO based first-principles
calculations on rather small systems (i.e. mostly less than 10 atoms).

However, numerical Coulomb has several advantages, besides being basis
set independent. In a Gaussian based approach, it can be particularly
efficient when dealing with heavier elements with a large basis set
that involves higher angular polarization functions. It can also be
combined with far-field multipole expansion naturally for a more efficient
computational scaling. Since each center is treated independently,
it also be parallelized easily for large scale calculations \cite{Franchini2013b,Franchini2014b}.

     Our interest in the basis-set-free numerical scheme stems 
from the limitation in using the standard analytic scheme for 
calculation of Coulomb potential of the scaled charge density 
needed in fully self-consistent implementation of the
locally scaled self-interaction-correction (LSIC) method developed by some of us.
The LSIC method identifies the one-electron self-interaction regions 
with the help of an iso-orbital indicator and determines the magnitude 
of SIC at each point according to the value of iso-orbital indicator \cite{zope2019lsic}.
The LSIC method which was earlier applied in a perturbative manner using the 
self-consistent Perdew-Zunger SIC (PZSIC) orbitals, have shown remarkable 
 improvement over the well known PZSIC method for many properties \cite{zope2019lsic}.
 It is the first one-electron SIC method that provides the atomization of 
 energies of AE6 database that are more accurate than the Perdew-Burke-Erzerhof 
 generalized gradient approximation 
 without spoiling the accuracy of barrier heights. This work also showed that 
 major errors made  by the LSDA functional can be removed by removing 
 self-interaction 
 errors using LSIC like approach. The self-consistent implementation of LSIC 
 method is therefore important and the present work, which will allow calculation 
 of Coulomb potential of a charge density scaled by an isoorbital indicator 
 is, the first step towards the self-consistent LSIC.

The manuscript is organized as follows:  we first review two major methods \cite{Becke1988d,Delley1990b}.
Then we performed a series of tests along with some numerical tricks
to enhance the radial grid efficiency. By analyzing and comparing
the data for both methods we were able to identify the origin of the
strengths and weakness in both methods. Finally, we propose a hybrid
approach to further optimize the performance of the numerical Poisson
solver. 

\section{Method review\label{sec:method}}

\subsection{Mesh generation}

The numerical Poisson solver utilizes so called multi-center grid
\cite{Becke1988e}, or often collectively referred to as Becke mesh,
that was originally designed for 3D molecular integrals for functionals
of the form $F(\rho(r),\nabla\rho(r),\nabla^{2}\rho(r)$). To properly
described the cusps at the nuclear cores, the grid itself is constructed
as a superposition of multiple spherical integration grids where each
spherical grid is constructed by multiplying a radial quadrature onto
a spherical mesh to form concentric spherical shells of mesh centered
on each atomic site. 

While Lebedev quadrature is generally believed to be the most efficient
for spherical mesh \cite{lebedev1976quadratures}, there have been
a wide variety of options for choosing radial grids \cite{El-Sherbiny2004b,Gill1993b,Gill2003b,Kakhiani2009b,Krack1998b,Lindh2001b,Mura1996b,Murray1993b,Treutler1995b},
and each has its own advantages. In this study we adopt the radial
quadrature proposed by Mura and Knowles \cite{Mura1996b}, as it has
been proven to be numerically efficient \cite{Kakhiani2009b} and
has a simple form 

\begin{equation}
r_{i}=-\alpha\mathrm{ln}(1-x_{i}^{m})\label{eq:ri}
\end{equation}
where 
\begin{equation}
x_{i}=\frac{i}{N+1},\;i=1,2,...N\label{eq:xi}
\end{equation}
N is the number of radial points, and both $\alpha$ and $m$ are empirical parameters controlling how the points are distributed.
The recommended values are $m=3$ and $\alpha=5.0$ (7.0 for alkali and rare-earth metals) \cite{Lindh2001b}.

This simple expression makes it easier for analytically evaluating the coefficients
of the FDM operator for solving 1D radial Poisson equations. 

\subsection{Partition functions }

The overlap between the meshes from different centers can be avoided
by scaling down the integrated function $\rho(r)$ with the partition
weight function $w_{n}$ as 

\begin{equation}
\rho_{n}(r)=\rho(r)w_{n}(r)\label{eq:rho_wn}
\end{equation}
and 

\begin{equation}
\rho(r)=\sum_{n}\rho(r)w_{n}(r).\label{eq:sum_rho_wn}
\end{equation}

Below we refer to this process as single-centerization where the whole
space is partitioned into multiple independent cells, similar to the
Voronoi cells but with smoother transition boundaries and each cell
contains only one nucleus.

The relative weight functions $w_{n}$ are constrained to satisfy
the condition

\begin{equation}
\sum_{n}w_{n}=1\label{eq:1wn}
\end{equation}
and defined as 

\begin{equation}
w_{n}(r)=\frac{\mathfrak{\mathcal{P}}_{n}(r)}{\mathfrak{\mathcal{\sum_{\mathrm{m=1}}^{\mathrm{N_{atoms}}}P}}_{m}(r)}\label{eq:wn_def}
\end{equation}
where $\mathfrak{\mathcal{P}}_{n}(r)$ is the cell function. 

Over the years, many cell function generating schemes have been
proposed. 
Begins with the original one proposed by Becke \cite{Becke1988e}
that requires predetermined empirical parameters to accommodate different atomic species of different sizes. 
Delley \cite{Delley1990b} provides some easier atomic density
dependent implementations to naturally account for the molecules that
contain different atomic species. Some weight schemes are also proposed
to achieved better scaling for finite systems \cite{Stratmann1996b}
and for periodic systems \cite{Franchini2013b}. There are also variants
based on Becke's original scheme which are designed for certain properties
of interests \cite{Gharibnejad2021b,Laqua2018b}.

Although the mesh generated with these cell function schemes can 
be applied to both general 3D molecular integration and numerical Poisson
solver, they are not necessary the same \cite{Delley1990b}.
The weight functions constructed with different cell functions are generally
in good agreement, however our experience suggests that the scheme
proposed by Stratmann \cite{Stratmann1996b} gives the most accurate
results in some geometries which confirms the earlier
report \cite{Blum2009b}. Therefore the Stratmann's approach is adopted
for all the calculations presented in this current study.

\subsection{Single centerization and multipolar expansion}

The starting point is to replace the source term in Poisson equation
with a single-centerized charge defined in Eq. \ref{eq:rho_wn}. This 
results in significant simplification as one only needs to  deal
with the equation containing one atomic center at a time.

\begin{equation}
\nabla V_{n}(r)=-4\pi\rho_{n}(r).\label{eq:poisson_n}
\end{equation}
As the mesh for each center is constructed based on spherical mesh,
we can efficiently expand both $\rho_{n}$ and $V_{n}$ with
spherical harmonic functions \cite{Jackson:100964} respectively as
\begin{equation}
\rho_{n}(r,\theta,\phi)=\sum_{lm}\rho_{lm}^{(n)}(r)Y_{lm}(\theta,\phi)\label{eq:rho_lm}
\end{equation}
and

\begin{equation}
V^{(n)}(r=|r'-r_{n}|,\theta,\phi)=\sum_{lm}^{l_{max}}V_{lm}(r)Y_{lm}(\theta,\phi),\label{eq:V_n_sh}
\end{equation}
where 

\begin{equation}
\rho_{lm}^{(n)}(r)=\intop_{\Omega}\rho_{n}(r,\theta,\phi)Y_{lm}(\theta,\phi)d\Omega\label{eq:rho_lm_n}
\end{equation}
and $V_{lm}^{(n)}(r)$ are to be solved.

\subsection{Radial Poisson }

For each {\it lm} component in SHE, the angular and radial degrees of freedom
are separable, the Poisson equation can then be converted into a set
of 1D problems. There are two major approaches to solve for $V_{lm}$
which is where Delley and Becke's methods differ. 

In Becke's approach, $V_{lm}(r)$ are obtained by solving a set of
1D differential equations. With the substitution 
\[
V_{lm}(r)=\frac{U_{lm}(r)}{r},
\]
one arrives at the following general expression 
\begin{equation}
\frac{\partial^{2}}{\partial x^{2}}U_{lm}(r)+p(r)\frac{\partial^{2}}{\partial x^{2}}U_{lm}(r)-q(r)U_{lm}(r)=f(r),\label{eq:poisson_rad}
\end{equation}
where
\begin{equation}
x_{i}=(1-\mathrm{exp}(-r_{i}/\alpha))^{\frac{1}{m}}.\label{eq:xi_r}
\end{equation}

Eq. \ref{eq:poisson_rad} can be solved using FDM with the coefficients
calculated through

\begin{equation}
p(r)=\frac{\partial^{2}x}{\partial r^{2}}/\left(\frac{\partial x}{\partial r}\right)^{2},\label{eq:p_r}
\end{equation}

\begin{equation}
q(r)=\frac{l(l+1)}{r^{2}\left(\frac{\partial x}{\partial r}\right)^{2}}\label{eq:q_r}
\end{equation}
and

\begin{equation}
f(r)=-4\pi\rho_{lm}(r)/\left(\frac{\partial x}{\partial r}\right)^{2}.\label{eq:f_r}
\end{equation}
The boundary conditions (BCs) are imposed to be
$U_{lm}(r)=0$ for $r$=0 and $r\rightarrow\infty$ for
all $lm$'s except for the monopole $l=0$ where 
$U_{00}(r\rightarrow\infty)=\sqrt{4\pi}q_{n}$.

On the other hand, Delley applied the GF solution to the Laplacian
\cite{garfken67:math} to integrate with charge directly for $I_{lm}(r_{i})$
\cite{Delley1990b,Franchini2014b,Blum2009b} in the following form

\begin{equation}
I_{lm}(r_{i})=\frac{1}{r_{i}^{l+1}}\intop_{0}^{r_{i}}r'^{l+2}s_{lm}^{i}(r')dr'+r_{i}^{l}\intop_{r_{i}}^{\infty}\frac{s_{lm}^{i}(r')}{r'^{l-1}}dr'.\label{eq:I_j}
\end{equation}
where $s_{lm}^i$ is the fitted charge density evaluated at mesh point $r_i$.

In this study, we also try to further smooth out the fitted curve $s_{lm}^{i}(r')$
by rearranging the integral form of both terms in Eq. \ref{eq:I_j}
into the following expression
\begin{equation}
I_{lm}(r_{i})=\frac{1}{r_{i}^{l+1}}\intop_{0}^{r_{i}}r'^{l+2-p}\left[r^{p}s_{lm}^{i}(r')\right]dr'+r_{i}^{l}\intop_{r_{i}}^{\infty}\frac{\left[r^{p}s_{lm}^{i}(r')\right]}{r'^{l-1+p}}dr'\label{eq:I_j_2}
\end{equation}  
where $p=0,1$ or $2$ and we perform cubic spline interpolation
on $\left[r^{p}s_{lm}^{i}(r')\right]$ instead. The solution of the
radial Poisson defined in Eq. \ref{eq:V_n_sh} can be expressed as 

\[
V_{lm}(r)=\frac{4\pi}{2l+1}I_{lm}(r).
\]

This approach is in principle straightforward, however the integration
for Eq. \ref{eq:I_j} and \ref{eq:I_j_2} is tricky, as the integration is done for every $r_{i}$ 
and can no longer benefit from the quadrature weights. The numerical
instability could potentially occur in both the interpolation of the
charge and the evaluation of Eq. \ref{eq:I_j} as the term involves
higher order polynomial terms. In some of the earlier works such as
Ref. \cite{Blum2009b}, the integration is first
evaluated by spline-interpolating the multipolar charge density onto
a denser mesh and then the integration is done numerically. Here,
we adopt the most recent Franchini's integration scheme \cite{Franchini2014b}
instead where the multipolar charge density, $s_{lm}^{i}(r)$ at interval
$[i,i+1]$, is expressed as piece-wise polynomials

\begin{equation}
s_{lm}^{j}(r)=a_{j}+b_{j}r+c_{j}r^{2}+d_{j}r^{3}\label{eq:sjlm}
\end{equation}
then the integrals in Eq. \ref{eq:I_j} can be evaluated analytically
for each segment. This approach should, in principle, yield the most
accurate results for a given spline compared to the former and is
also computationally more efficient.

\subsection{Reconstructing the Coulomb potential}

Once the 1D radial solutions $V_{lm}$ are calculated, the total Coulomb
potential due to the charge in a particular cell can be reconstructed
on any integration mesh using Eq. \ref{eq:V_n_sh}, where $V_{lm}=r^{-1}U_{lm}(r_{i})$
for Becke's FDM and $V_{lm}=\frac{4\pi}{2l+1}I_{lm}(r_{i})$ for Delley's
GF integral. For $V_{lm}$ of any arbitrary given $r'$ that does not
coincide any radial grid point $r_{i}$, the evaluation is done using
cubic spline interpolation

\begin{equation}
V^{(n)}(r_{i}=|r'-r_{n}|,\theta_{i},\phi_{i})=\sum_{lm}^{l_{max}}V_{lm}^{interp}(r_{i},\theta_{i},\phi_{i}).\label{eq:V_n}
\end{equation}

\subsection{Screening charge }

As the charge density $\rho(r)$ normally possesses rapid changing peaks
in the proximity of nuclei, it is often useful to introduce some type
of screening charge \cite{Delley1990b,Blum2009b}.
The $\rho_{\mathrm{screen}}$ is usually chosen to smooth out the
curve around the cusps in $\rho(r)$ and that its corresponding Coulomb
potential can be evaluated analytically and efficiently. In which
case we only need to consider the Poisson problem due to the variation
from the screening charge. A convenient option is the superposition
of spherically symmetric neutral atomic charge densities \cite{Blum2009b,Delley1990b}.

\begin{equation}
\Delta\rho(r)=\rho(r)-\sum_{sites}\rho_{\mathrm{atoms}}(r),\label{eq:rho_diff}
\end{equation}
and the final total Coulomb potential can be restored by adding the
analytically evaluated potential $V_{\mathrm{atoms}}$ due to the
screening charge,

\begin{equation}
V=\Delta V+V_{\mathrm{atoms}}.\label{eq:rho_tot}
\end{equation}

\section{computational details}
We use Pederson-Porezag basis set \cite{Porezag_bsis} in all our calculations.
For the integration mesh, and for the numerical Poisson solver, Lebedev
unit sphere quadrature of order $l=47$, which contains 590 points,
is used. For FDM, instead of 7-point suggested by Becke and Dickson \cite{Becke1988d}, we use 11-point central difference formula for the middle
radial points and non-central difference for the points involve the
boundary to preserve the banded shape of the operator matrix. 
However we observed that there is no noticeable difference beyond 7-point.
The coefficients for FDM operators are generated using Matlab function
\cite{FDMweight}. For the GF approach of solving radial Coulomb
potentials, we adopted Franchini's density fitting \cite{Franchini2014b}
with a slight modification as introduced in Eq. \ref{eq:I_j_2}. For
the interpolation of $V_{lm}$ onto any given radial point $r$, we
apply piece-wise cubic spline interpolation subroutine modified from
the subroutines in numerical recipes \cite{nrf77}.

As the treatment for the angular part of the solutions is well-defined within SHE,
the main purpose of this study will be focusing on improving the radial solutions, 
we choose a generic set of parameters $\alpha=6$
and $m=3$, similar to the recommended values in Ref. \cite{Mura1996b},
for the radial quadrature in all our calculations unless explicitly
stated.

\section{Analysis and discussion}

\subsection{Atomic system (single-center)}

\begin{figure}
\centerline{\includegraphics[scale=0.66]{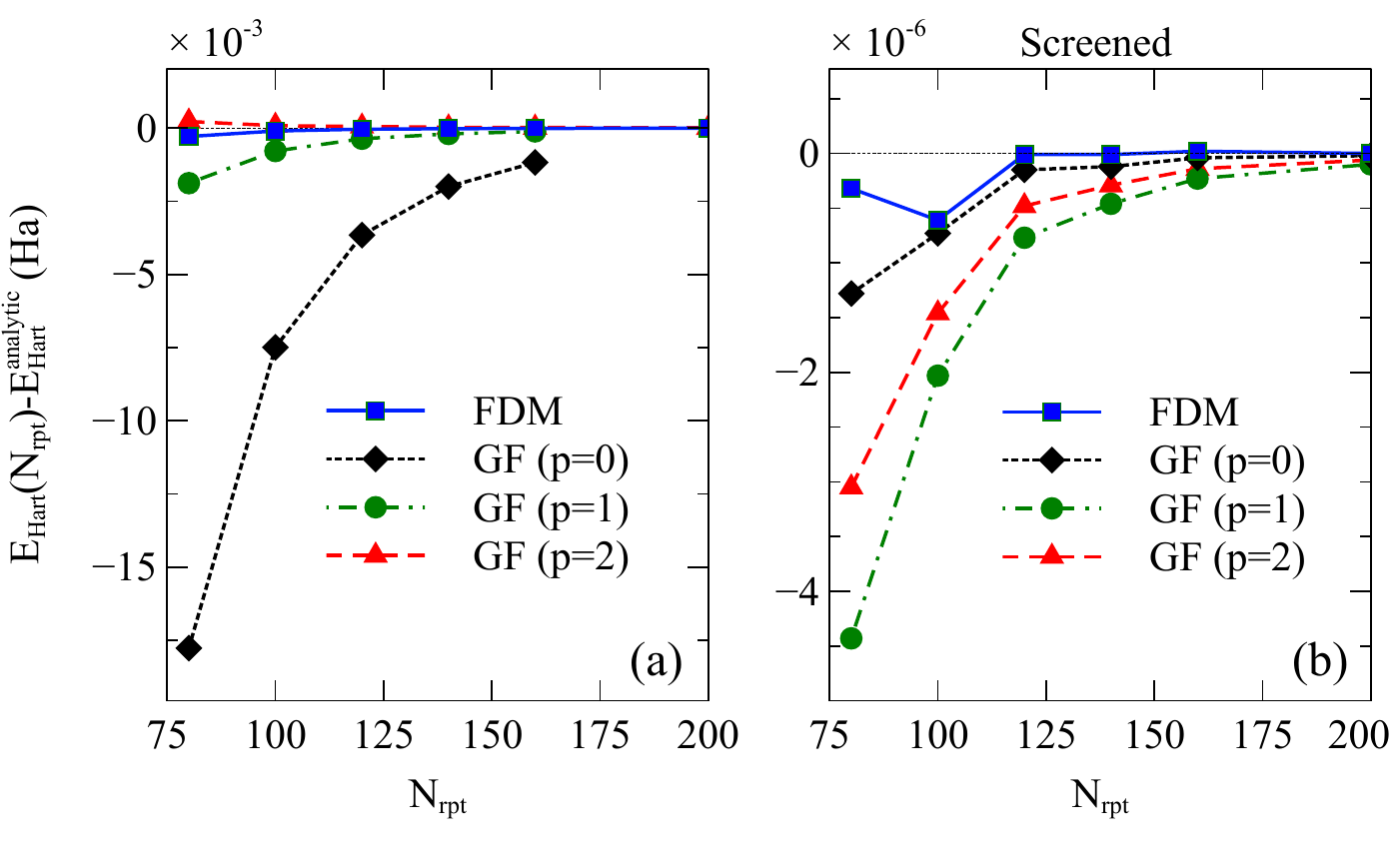}}

\caption{\label{fig:Mn_atm}Hartree energy $E_{\mathrm{Hart}}$ for a single
Mn atom is calculated using both FDM and GF (a) without and (b) with
screening charge. Different modifications of charge interpolation
are also performed as a comparison as shown for $p=0,1$ and $2$. }
\end{figure}

We begin with the analysis using both methods on single atom systems.
As the space partitioning is not required, we get a better grasp of
how efficiently the methods perform with respect to a given radial
quadrature. Fig. \ref{fig:Mn_atm} shows the error of Hartree energy
for Mn atom (a) without and (b) with charge screening compared to
the reference value, where the Coulomb potential is analytically evaluated \cite{Termath1994b}.

From Fig. \ref{fig:Mn_atm}(a), in their primitive form, FDM shows
a clear advantage over GF ($p=0$), as it requires less than 120 radial
points to reach $10^{-5}$ Hartree accuracy while GF provides
barely a mHatree accuracy. However, once we have introduce
the additional polynomial
$r^p$ ($p=1$ and $2$) as defined in Eq. \ref{eq:I_j} to
smooth out the density peak close to the origin, the performance 
of the GF improves  drastically.
Particularly with $p=2$, the GF result
is almost comparable with FDM. This suggest that the charge density
variation appears to be too large for the cubic spline polynomials
to describe the nucleus region as efficiently, and smoothing out the
curve reduces the interpolation error.

Fig. \ref{fig:Mn_atm}(b) shows the results when screening charge
is applied. Without surprise, with the inclusion of screening charge
that naturally reduces the curvature
of charge density near the nucleus, the error for both
methods are greatly reduced to about $10^{-7}$ Hartree, with only
about 100 points. It is obvious that FDM still outperforms all the
GF cases with different $p$'s, even though the accuracy gap has become
much smaller. It is also interesting to note that in this particular
case, $p=1$ or $2$ does not seem to improve the result at all. 

\begin{figure}
\centerline{\includegraphics[scale=0.66]{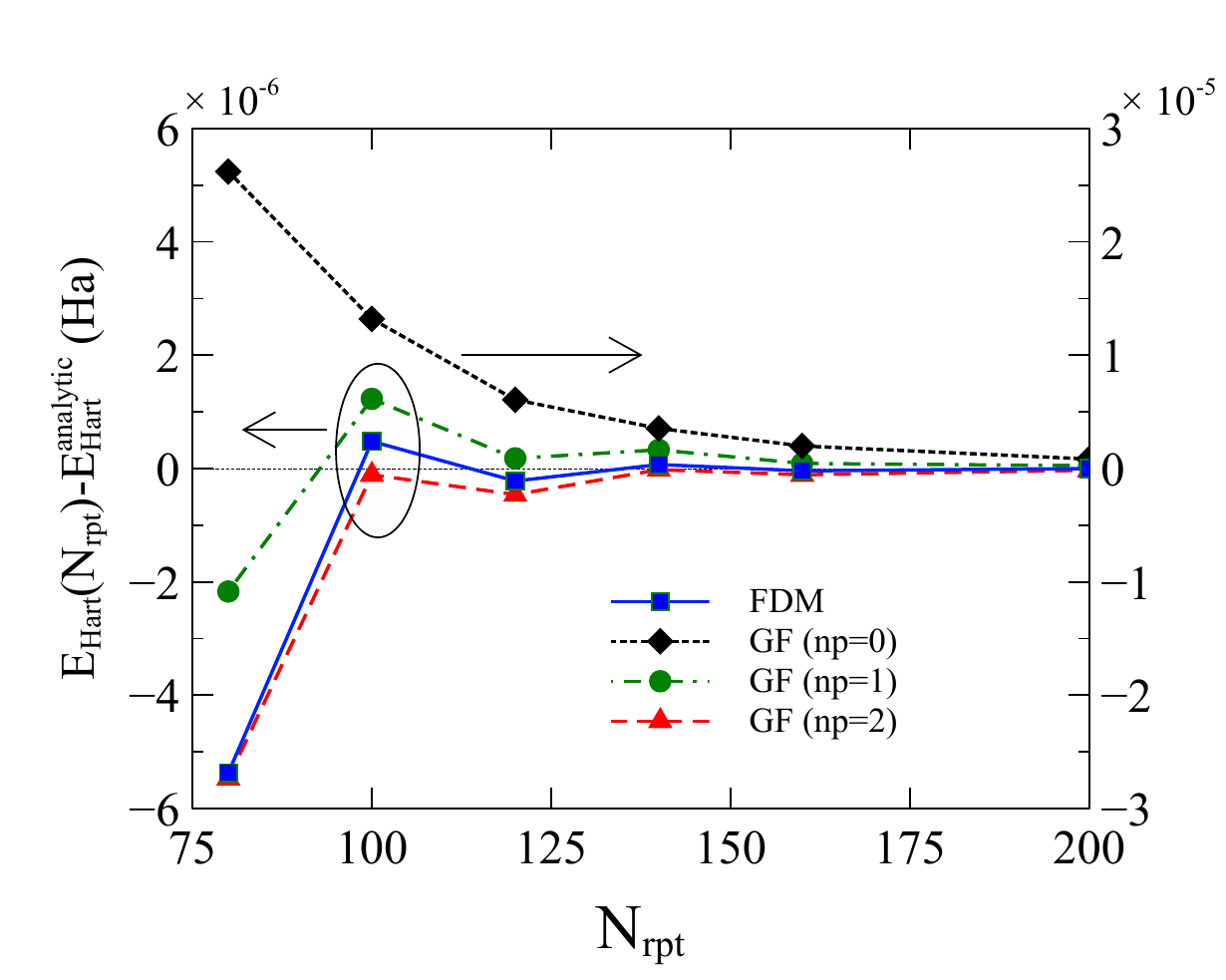}}

\caption{\label{fig:Cu_atm}The error of Hartree energy for a single Cu atom
versus number of radial points $N_{rpt}$ of modified charge interpolations
(i.e. $r^ps_{lm}^{i}$ with different 
shown. Generally, the best performance can be achieved with $n=2$. }
\end{figure}

Fig. \ref{fig:Cu_atm} presents the same analysis as Fig. \ref{fig:Mn_atm}
(b) but on a single Copper atom. Please note that the black dashed
line for GF ($p=0$) follows the axis to the right of a larger energy
scale, while the other three follow the left axis. By applying screening
charge, all four cases offer adequate accuracy with 100 radial points
(i.e. $\apprle10^{-5}$ Hartree). Noticeably, the error of FDM is
still smaller than all three GF cases, while the result of GF improves
as $p$ increases, especially when compared to its original form ($p=0$)
where the error is at least an order of magnitude larger than the
rest. 

In this case, opposite to the Mn atom, the polynomial factor improves
the accuracy even in the presence of screening charge. Although both
screening charge and $r^p$ can improve the interpolation, it is
still possible that the piece-wise cubic spline is not sufficient,
as it incorporates only two adjacent data points (i.e. radial quadrature)
for each segment, while FDM naturally incorporates multiple data points
during the 2N+1-point ($N=5$) difference operator construction.
As a result, FDM is always more accurate with a more stable convergence
even without screening charge. 

In principle, the result for GF can be further improved by optimizing
radial parameters $\alpha$ and $m$. However, the purpose of this
analysis is to demonstrate the effectiveness of the FDM, where the
method generally describe the monopole better, and is less sensitive to
the radial quadrature parameters.  This is particularly useful for solving for the 
potentials due to orbital densities needed in the self-consistent one electron self-interaction
methods \cite{perdew1981pzsic,vydrov2006osic,yamamoto2020sosic,zope2019lsic} where  suitable screening charges are
either  difficult or expensive to find. 

\subsection{ Molecular system (multi-center)}

\begin{table}
\fontsize{8}{8}\selectfont
\def\arraystretch{2.0}
\begin{tabular}{c|r@{\extracolsep{0pt}.}l|r@{\extracolsep{0pt}.}l|r@{\extracolsep{0pt}.}l|r@{\extracolsep{0pt}.}l}
 & \multicolumn{2}{c|}{$E_{Hart}$} & \multicolumn{2}{c}{} & \multicolumn{2}{c}{$\Delta E$} & \multicolumn{2}{c}{}\tabularnewline
\cline{2-9} \cline{4-9} \cline{6-9} \cline{8-9} 
Molecule & \multicolumn{2}{c|}{Analytic} & \multicolumn{2}{c|}{FDM} & \multicolumn{2}{c|}{GF} & \multicolumn{2}{c}{Hybrid}\tabularnewline
\hline 
Glycine(10) & 549&790197 & -2&18E-04 & 8&30E-06 & -4&98E-06\tabularnewline
Benzene(12) & 311&939018 & -2&71E-04 & 1&40E-05 & 5&01E-06\tabularnewline
Decanol(33) & 816&263155 & -3&54E-04 & 1&00E-05 & 3&47E-06\tabularnewline
Pentacene(36) & 1892&400793 & -8&97E-04 & 2&53E-05 & 8&54E-06\tabularnewline
C$_{60}$(60) & 9436&4919523 &   -1&69E-03 &  5&45E-05 & 1&28E-05\tabularnewline
\end{tabular}\caption{\label{tab:error-mol}Error/atom for a set of testing molecules of
different sizes range from 10 to 60 atoms}
\end{table}

\begin{figure}
\centerline{\includegraphics[scale=0.5]{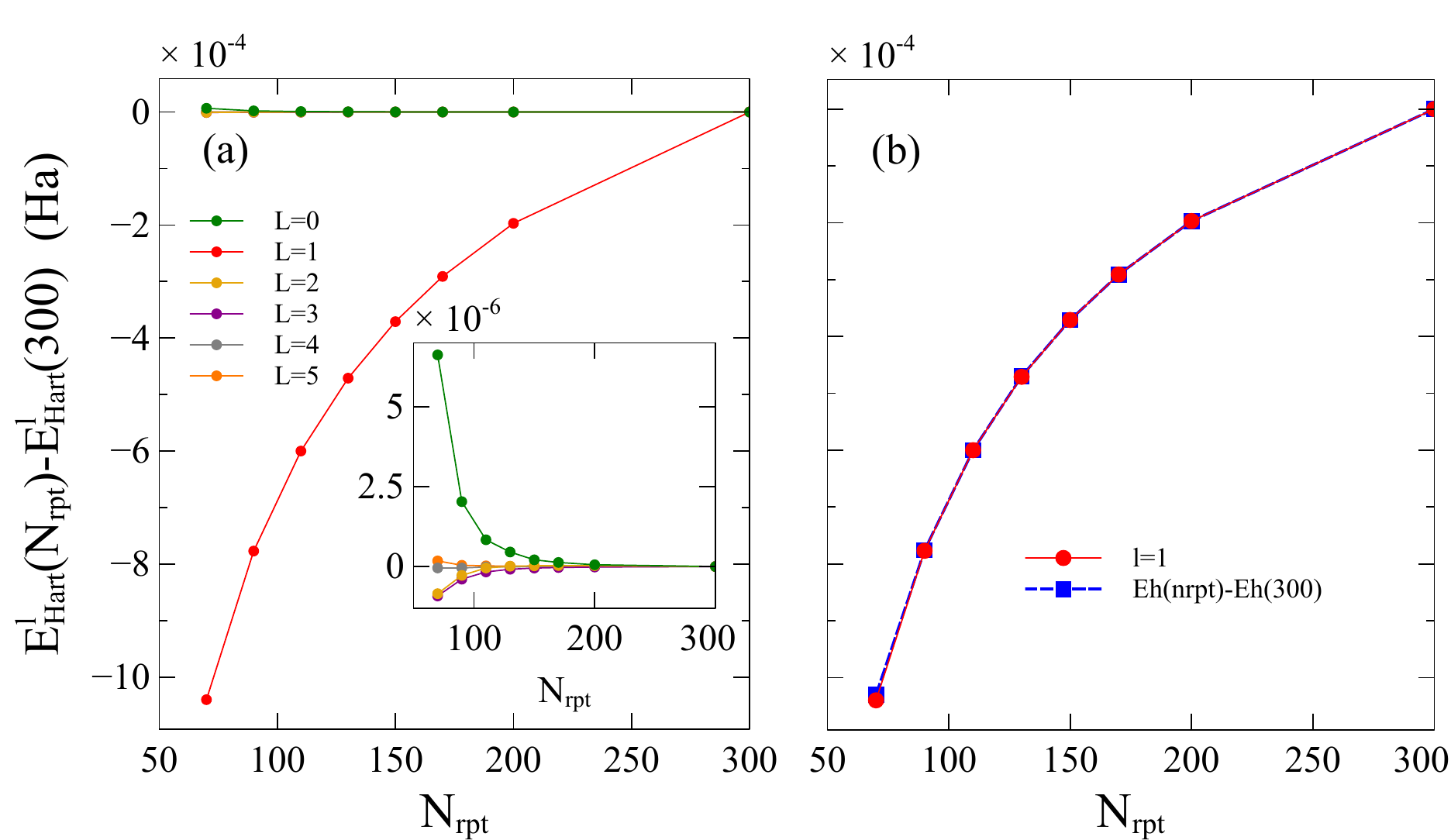}}

\caption{\label{fig:E_l}The deviation of Hartree energy of Glycine molecule
of each l-component from the the same quantity but evaluated with $N_{rpt}=300$ as
a convergence test. (a) The Hartree energy contributions from the
lowest six orders $(l=0\sim5$). (b) the largest error contribution
($l=1$) compared to the error of the total Hartree energy. }
\end{figure}

\begin{figure*}
\centerline{\includegraphics[scale=0.65]{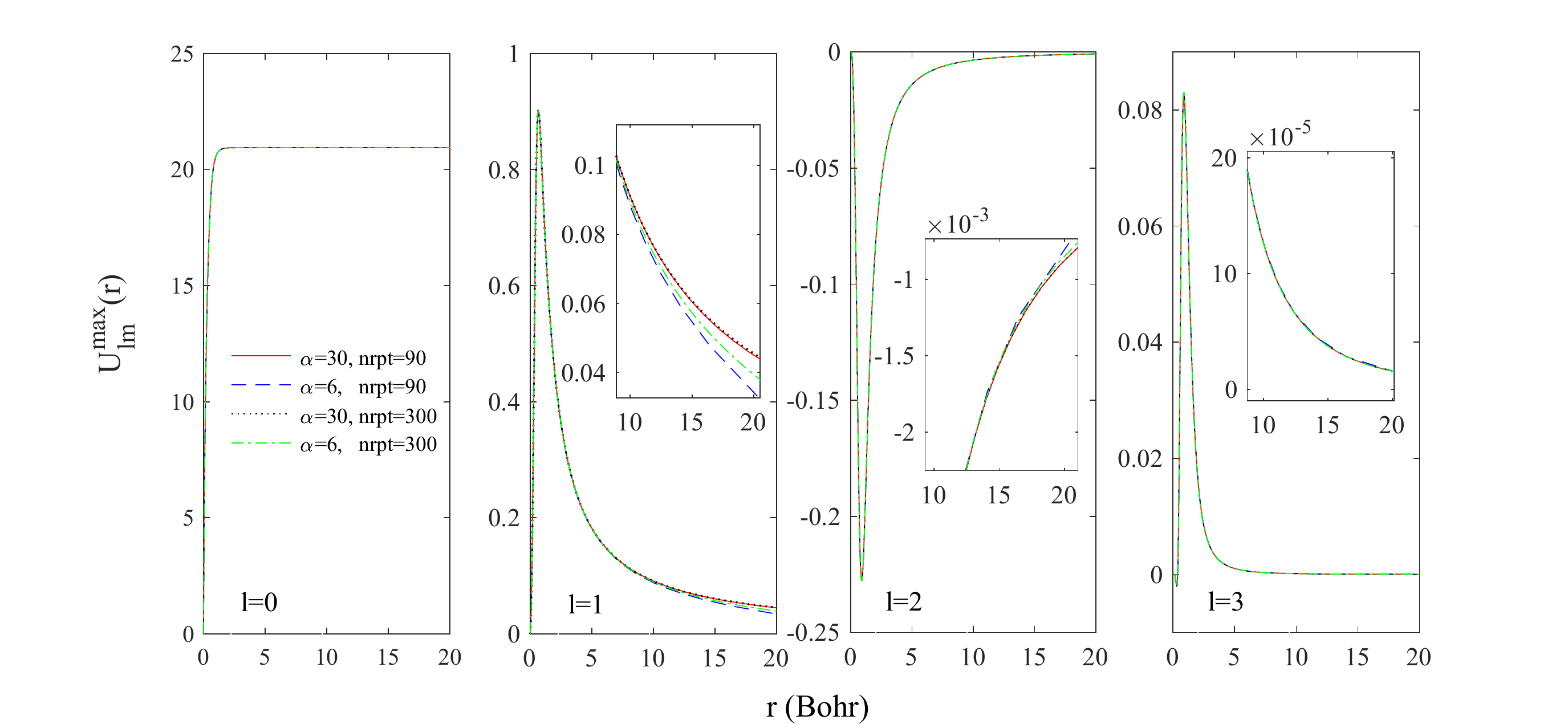}}

\caption{\label{fig:ulm_r}Four sets of radial parameters are used to test
the convergence of the radial solutions. The maximum component of
the radial solutions ($U_{lm}^{max}$) of the four lowest expansion
orders (up to $l=3$) for the oxygen site in Glycine molecule are
shown as (a)-(d). For (b)-(d) the insets present a closer look of
the convergence of the tails. }
 
\end{figure*}

Although FDM appears to be very efficient for the single atom case
that contains only one center, the situation reversed in the multi-center
systems (i.e. molecules). Table \ref{tab:error-mol} shows error
per atom for different radial Poisson solvers on several selected molecules
ranging from 10 to 33 atoms. One can see that the FDM becomes particularly
problematic. Not only is the error generally large, it also scales
with the system size. As previous studies have already pointed out
\cite{Termath1994b}, the error appears to be accumulative and becomes
worse with increasing system size. The increasing error with system 
size has limited the method from being used widely.

Furthermore, we also noticed that the result is very sensitive to
the radial mesh scaling factor $\alpha$. An unreasonably large $\alpha$
often yields better accuracy than a small one with larger $N_{rpt}$.
This is rather counter-intuitive. Since FDM performs so well in the single atom systems 
which means the solver is capable of describing the monopole that contains the sharpest peak. 
Naively, one would expect the charge partitioning should mostly affect the multipolar expansion (i.e. requires larger $l$).

\begin{table}
\fontsize{8}{8}\selectfont
\def\arraystretch{2.0}%

\subfloat[\label{tab:tab_fdm}FDM]{%
\begin{tabular}{l|cccccc}
$N_{rpt}$ & $l=0$ & $l=1$ & $l=2$ & $l=3$ & $l=4$ & $l=5$\tabularnewline
\hline 
70 & 1.84E-05  & -1.38E-03 & -3.96E-07 & -2.98E-06 & -1.42E-06 & -4.18E-07\tabularnewline
90 & 6.75E-06  & -1.03E-03 & -9.24E-08 & -1.10E-06 & -5.13E-07 & -1.64E-07\tabularnewline
110 & 2.38E-06 & -8.00E-04 & -5.00E-08 & -4.97E-07 & -2.34E-07 & -7.71E-08\tabularnewline
130 & 1.21E-06 & -6.27E-04 & -1.97E-08 & -2.47E-07 & -1.18E-07 & -3.81E-08\tabularnewline
150 & 6.80E-07 & -4.94E-04 & -1.60E-08 & -1.31E-07 & -6.07E-08 & -1.86E-08\tabularnewline
170 & 4.19E-07 & -3.88E-04 & -3.55E-09 & -8.36E-08 & -3.77E-08 & -1.18E-08\tabularnewline
200 & 2.09E-07 & -2.62E-04 & -1.45E-09 & -3.99E-08 & -1.78E-08 & -5.90E-09\tabularnewline
\end{tabular}

}

\subfloat[GF ($p=0$)]{%
\begin{tabular}{c|cccccc}
$N_{rpt}$ & $l=0$ & $l=1$ & $l=2$ & $l=3$ & $l=4$ & $l=5$\tabularnewline
\hline 
70 & -1.24E-03 & -1.04E-04 & -1.20E-05 & -3.64E-06 & -2.10E-06 & 1.03E-06\tabularnewline
90 & -4.63E-04 & -3.84E-05 & -4.38E-06 & -1.42E-06 & -8.32E-07 & 3.11E-07\tabularnewline
110 & -2.09E-04 & -1.72E-05 & -1.98E-06 & -6.64E-07 & -3.93E-07 & 1.18E-07\tabularnewline
130 & -1.06E-04 & -8.72E-06 & -9.94E-07 & -3.35E-07 & -2.05E-07 & 5.31E-08\tabularnewline
150 & -5.84E-05 & -4.81E-06 & -5.47E-07 & -1.77E-07 & -1.10E-07 & 2.95E-08\tabularnewline
170 & -3.40E-05 & -2.79E-06 & -3.20E-07 & -1.13E-07 & -6.60E-08 & 1.72E-08\tabularnewline
200 & -1.59E-05 & -1.31E-06 & -1.50E-07 & -5.21E-08 & -3.20E-08 & 7.17E-09\tabularnewline
\end{tabular}
}

\caption{\label{tab:Eh_l} The $l$-resolved energy convergence for the Glycine molecule of 
(a) FDM and (b) GF method versus number of radial points.}
\end{table}

Table \ref{tab:Eh_l} shows the convergence of the $l$-resolved Hartree energy 
contributions for Glycine defined as
$\Delta E_{\mathrm{Hart}}^{l}(N_{rpt})=E_{\mathrm{Hart}}^{l}(N_{rpt})-E_{\mathrm{Hart}}^{l}(300)$, 
where

\[
E_{\mathrm{Hart}}^{l}=\frac{1}{2}\int\rho(r)V_{n}^{l}(r)dr
\]
is evaluated from the potential defined as

\[
V_{n}^{l}(r)=\sum_{m=-l}^{l}V_{lm}(r)Y_{lm}(\theta,\phi).
\]

The data reveal something rather intriguing. Particularly in FDM,
a perfect convergence trend can be seen in all terms except for $l=1$
which exhibits at least $3\sim4$ orders of magnitude larger error
than other $l$ components. The result in Table \ref{tab:tab_fdm} for
FDM is also plotted in Fig. \ref{fig:E_l} for better visualization.
From Fig. \ref{fig:E_l}(a), it becomes obvious that the error is
entirely coming from $l=1$ while all other terms converge rapidly
to $0$, even the dominant $l=0$ term is well converged around $N_{rpt}$=100.
Fig. \ref{fig:E_l}(b) compares the error only from $l=1$ and the
error of the entire Hartree energy. The two curves coincide nearly
perfectly which further confirms not only that $l=1$ is the only
main source of error but also that the contributions from all the
higher orders terms converge perfectly.

To better understand this rather peculiar cause of error, we analyze the radial
part of the potential, $U_{lm}$, for the Oxygen site in the Glycine molecule 
of the lowest
four orders as shown in Fig. \ref{fig:ulm_r} using four very different
sets of radial mesh parameters. Since not all terms are nonzero due
to symmetry consideration, only the largest contribution is shown
for every $l$. The parameters $(\alpha,N_{rpt})$ for each of the four
quadratures are chosen as $(6.0,90)$, $(6.0,300)$, $(30.0,90)$ and $(30.0/300)$.
A large number radial points $N_{rpt}=300$, is chosen here intended
as an accurate reference. A large scaling factor $\alpha$ simply
stretches out the radial mesh to cover wider space range and as a
result will make the mesh grid sparse.  At the first glance, all four radial
meshes coincide nicely at $l=0$ and $l=4$ (and the
same for all the higher order terms not shown here). On
the other hand pronounced deviation in $l=1$ and a very small but noticeable
difference in $l=2$ can be seen, which indicates that these two l-orders
are more sensitive to the radial grid. Upon closer inspection, only
the ones with the larger $\alpha$ have converged for $l=1$ regardless
what $N_{rpt}$ is. This means that $N_{rpt}=90$ is sufficient to describe
the region around nucleus and the maximum range of the quadrature $r_{max}=r_{N_{rpt}}$,
scaled directly by $\alpha$ ( see Eq. \ref{eq:ri}) has a significant  influence on the accuracy.

Without surprise, $(6,300)$ gives a better result than $(6,90)$, but
this is partly because the increase of $N_{rpt}$ also increases the
$r_{i}^{max}$(=$20.5395$ for $(6,90)$ and =$27.6709$ for $(6,300)$,
rather than the fineness of the mesh. It is now clear that the
source of discrepancy
is about how the asymptotic behavior is described by FDM rather than the
description of the nucleus core region. 
By looking at the asymptotic tails at a larger distance $r=20$, $U_{lm}^{max}$ for both
$l=0$ and $l=4$ (and above) have reached their asymptotic limit
$\sqrt{4\pi}q_{n}$ and $0$ respectively. On the other hand for $l=1$, $U_{1m}^{max}(r=20)$
is still large and decreasing to $0$ in an extremely
slow pace (roughly $1/r$), hence induces a large deviation. For the
$l=2$ term, there is a visible but much smaller deviation at $r=20$,
therefore leaves smaller room for discrepancy compare to $l=1$. 

Intuitively speaking, if $U_{lm}(r)$ at the largest radial distance
$r=r_{N_{rpt}}$ has not yet approached close enough to the asymptotic
value (i.e. $\sqrt{4\pi}q_{n}$ for $l=0$ or $0$ for $l\neq0$),
a significant error is expected, as FDM only knows the boundary condition
at the infinity, the behavior beyond the last radial point $r_{N_{rpt}}$
can only be extrapolated from the last few radial points. This explains
why a significant error only emerges in the lower order terms but
not $l=0$, as it reaches $\sqrt{4\pi}q_{n}$ rapidly within merely
few Bohr's, while higher order terms are less likely to suffer from
this since they are shorter in range and decay to $0$ much faster.

This also explains why the method still works well for smaller molecules, 
as the deviation in the $l=1$ term in Fig. \ref{fig:ulm_r} is not obvious until $r$ is large 
enough. It is also worth mentioning that in the original attempt, Becke uses Gauss\textendash Chebyshev
formula of the second kind for the radial quadrature, which is known
to emphasize too much in the extended region while not putting enough
points into the chemical bonding region \cite{Kakhiani2009b}. However,
despite not being integrationally optimal, using this quadrature could potentially
mitigate the boundary issue appears in $U_{1m}$ terms described above, particularly for for smaller size systems.

\subsection{Hybrid radial solver}

\begin{figure}
\includegraphics[scale=0.55]{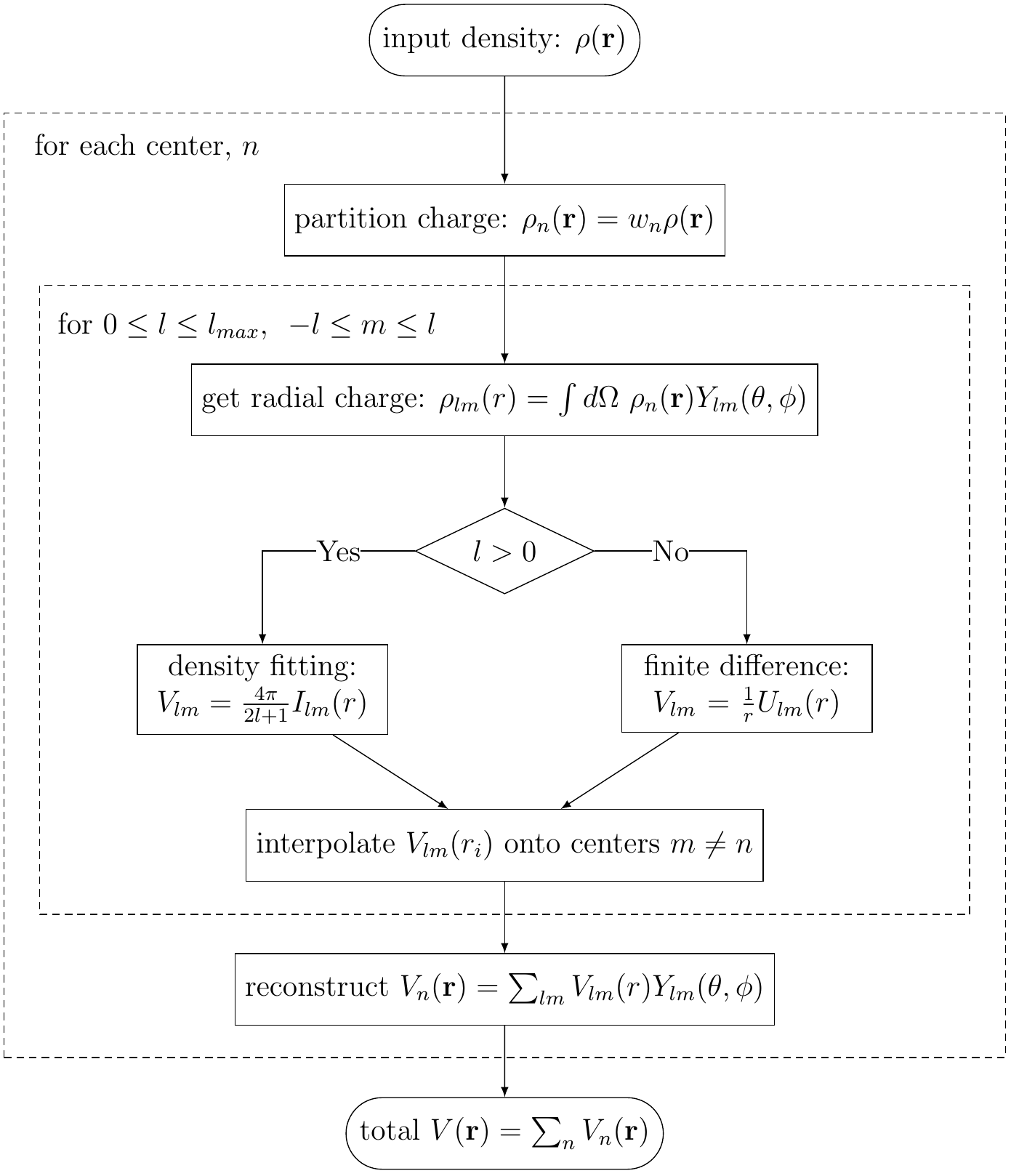}

\caption{\label{fig:flow}The flow chart of the numerical Poisson solver.}
\end{figure}

 As discussed in the previous section, the FDM method struggles 
for $l=1$ and possibly $l=2$. On the other hand,
GF method only requires the evaluation of charge
integrals of the form $r^{k(l)}\rho_{lm}(r)$ (see Eq. $\ref{eq:I_j}$
and $\ref{eq:I_j_2}$), where $k$ is just some integer depends on
$l$ chosen manually and does not suffer from the same ``boundary'' issue.
GF generally converges rather nicely with the number of the radial
points for all terms except for $l=0$ as shown in the Table \ref{tab:Eh_l}.
This is because the charge density in each center is more localized
in space after the truncation imposed by the weight function. 

By considering the strengths of both methods, we propose a hybrid approach
where FDM is used for the ``near-field'' $l=0$ and the rest are
calculated using GF integration. The detail of the procedure is outlined
in Fig. \ref{fig:flow}. For a large system, the first term of Eq.
\ref{eq:I_j_2} which corresponds to the far-field multipole expansion,
is employed to ensure the asymptotic behavior is accurately preserved.
In the rest of the discussion, the screening charge is applied to
all the calculations unless explicitly stated.

The test used in analysis of FDM and GF method, is
also used to analyze the performance of the proposed hybrid approach.
The results are shown in Table \ref{tab:error-mol}. 
It is evident from the Table that using the same generic set of parameters
$(\alpha,N_{rpt})=(6.0,90)$, the accuracy of the proposed hybrid 
approach is improved by several times to an order of magnitude 
compared to  the GF method. More importantly,
the error per atom in the proposed hybrid approach
does not scale with the system size. 
According to our analysis, both FDM and GF describe higher order terms (i.e. l>3)
equally efficient and accurately. Therefore, one would expect to see the same improvement for FDM-based method as long as $l=1$ terms are calculated using GF.

\subsubsection*{Energy difference between different spin states}

\begin{table}

\fontsize{8}{8}\selectfont
\def\arraystretch{2.0}%


\begin{tabular}{cr@{\extracolsep{0pt}.}lccc}
\multirow{2}{*}{} & \multicolumn{3}{c}{p-C$_{8}$H$_{8}$} & \multicolumn{2}{c}{Cu$_{2}$Cl$_{6}$$^{2-}$}\tabularnewline
\cline{2-6} \cline{4-6} \cline{5-6} \cline{6-6} 
 & \multicolumn{2}{c}{$E_{\mathrm{analytic}}$} & $\Delta E$ & $E_{\mathrm{analytic}}$ & $\Delta E$\tabularnewline
\hline 
$E_{\mathrm{BS}}$ & -306&704861 & 1.19E-04 & -6028.542708 & -6.80E-06\tabularnewline
$E_{\mathrm{HS}}$ & -306&698685 & 1.22E-04 & -6028.541461 & -7.00E-06\tabularnewline
\hline 
$E_{\mathrm{BS}}-E_{\mathrm{HS}}$ & -0&006176 & -2.90E-06 & -0.001247 & 2.00E-07\tabularnewline
\hline 
Error (\%) & \multicolumn{2}{c}{} & -0.02\% &  & 0.01\%\tabularnewline
\end{tabular}

\caption{\label{tab:exchang}Total energy and the total energy difference between
high spin and broken symmetry states for p-C$_{8}$H$_{8}$, and Cu$_{2}$Cl$_{6}$$^{2-}$. 
All energies are in unit of Hartree}

\end{table}
\begin{table}

\fontsize{8}{8}\selectfont
\def\arraystretch{2.0}%

\begin{tabular}{cr@{\extracolsep{0pt}.}lccc}
\end{tabular}

\end{table}

As an assessment of accuracy of the proposed hybrid in practical DFT 
calculations, we compute the energy difference between different spin states,
namely high spin (HS) and broken symmetry (BS), for two systems p-C$_{8}$H$_{8}$
and Cu$_{2}$Cl$_{6}^{2-}$. The results are summarized in Table \ref{tab:exchang}. 

The error in the energy difference between different spin states in both cases is much smaller than the error in total energy alone by $1\sim2$ orders of magnitude, with the error percentage only about 0.01\%, as it is often easier to achieve higher accuracy in energy difference than in total energy alone. This makes the method particularly useful for studying properties involving energy difference such as exchange coupling or magnetic anisotropy energy (MAE). The similar pattern for the energy difference is also observed in Ref. \cite{Blum2009b}.



\begin{figure}
\centerline{\includegraphics[scale=0.66]{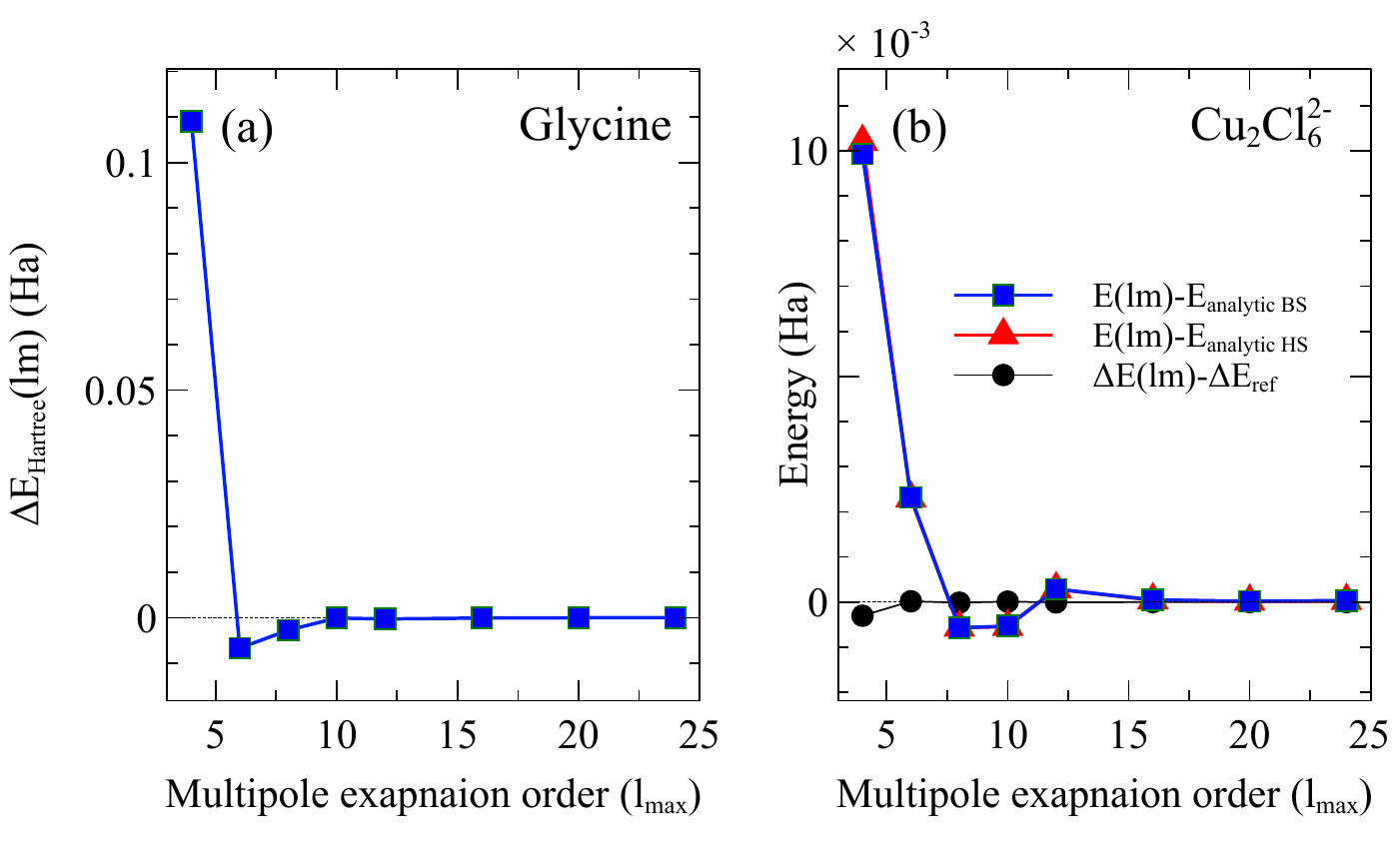}}

\caption{\label{fig:econv-l}Convergence versus multipolar expansion order
l for (a). the total energy error of Glycine compared to the reference
value and (b). the error of total energy of Cu$_{2}$Cl$_{6}^{2-}$
of both HS and BS spin states and of the energy difference between
the two. }
\end{figure}

An additional test for l-dependency is shown in Fig. \ref{fig:econv-l} for (a) the 
total energy of Glycine and (b) the total energy of Cu$_{2}$Cl$_{6}^{2-}$ of both 
HS and BS states and the energy difference between two states.
While all total energies requires around $l=16$ to reach convergence,    
it requires only up to $l=6$ (see Fig. \ref{fig:econv-l}(b)) to converge the energy difference. Similar patterns can also be found in many different contexts such as Brillouin zone integration in MAE. 


\section{Conclusions}

To summarize, we have reviewed two major numerical schemes (FDM and GF) for solving
Poisson equation. Both methods were analyzed with common numerical
tricks to improve the efficiency. We observed that, given the same set of radial parameters,
FDM appears to be more accurate describing the monopole ($l=0$) 
but it inherits a serious issue coming from
the multipolar contribution of mainly $l=1$ and sometimes $l=2$.
We propose a hybrid scheme by combining the two approaches, where the radial
potential of monopole (i.e. $U_{00}$) in spherical harmonic expansion
is calculated using FDM while the rest of the higher order terms are
calculated using GF's function integration. We then performed a series
of tests on different systems and the convergence with difference
numerical parameters to demonstrate effect of the implementation.
The overall results suggest that the accuracy is improved in all the
cases. More importantly our analysis show that like GF method,
the proposed hybrid 
approach is free from error accumulation with system size and is more
accurate for the monopole than the GF method. This observation 
and its favorable scaling especially when large basis sets are employed 
makes it an attractive method for studies on large complexes.

\section*{acknowledgments}
The authors thank Prof. Axel Becke  for the comments on the manuscript that 
helped in improving its clarity.  Authors gratefully acknowledge 
Dr. Carlos M. Diaz for discussions and assistance with parallelization of
the implementation. The computational resources from the Texas Advanced
Computing Center are gratefully acknowledged.
This work was supported by the US Department of Energy, Office of
Science, Office of Basic Energy Sciences, as part of the Computational
Chemical Sciences Program under Award No. DE-SC0018331. 

\section*{DATA AVAILABILITY}
The data that support the findings of this study are available from the corresponding author upon reasonable request




\bibliographystyle{apsrev4-1}
\bibliography{numerical_coulomb}

\begin{thebibliography}{53}%
\makeatletter
\providecommand \@ifxundefined [1]{%
 \@ifx{#1\undefined}
}%
\providecommand \@ifnum [1]{%
 \ifnum #1\expandafter \@firstoftwo
 \else \expandafter \@secondoftwo
 \fi
}%
\providecommand \@ifx [1]{%
 \ifx #1\expandafter \@firstoftwo
 \else \expandafter \@secondoftwo
 \fi
}%
\providecommand \natexlab [1]{#1}%
\providecommand \enquote  [1]{``#1''}%
\providecommand \bibnamefont  [1]{#1}%
\providecommand \bibfnamefont [1]{#1}%
\providecommand \citenamefont [1]{#1}%
\providecommand \href@noop [0]{\@secondoftwo}%
\providecommand \href [0]{\begingroup \@sanitize@url \@href}%
\providecommand \@href[1]{\@@startlink{#1}\@@href}%
\providecommand \@@href[1]{\endgroup#1\@@endlink}%
\providecommand \@sanitize@url [0]{\catcode `\\12\catcode `\$12\catcode
  `\&12\catcode `\#12\catcode `\^12\catcode `\_12\catcode `\%12\relax}%
\providecommand \@@startlink[1]{}%
\providecommand \@@endlink[0]{}%
\providecommand \url  [0]{\begingroup\@sanitize@url \@url }%
\providecommand \@url [1]{\endgroup\@href {#1}{\urlprefix }}%
\providecommand \urlprefix  [0]{URL }%
\providecommand \Eprint [0]{\href }%
\providecommand \doibase [0]{http://dx.doi.org/}%
\providecommand \selectlanguage [0]{\@gobble}%
\providecommand \bibinfo  [0]{\@secondoftwo}%
\providecommand \bibfield  [0]{\@secondoftwo}%
\providecommand \translation [1]{[#1]}%
\providecommand \BibitemOpen [0]{}%
\providecommand \bibitemStop [0]{}%
\providecommand \bibitemNoStop [0]{.\EOS\space}%
\providecommand \EOS [0]{\spacefactor3000\relax}%
\providecommand \BibitemShut  [1]{\csname bibitem#1\endcsname}%
\let\auto@bib@innerbib\@empty
\bibitem [{\citenamefont {Helgaker}\ \emph {et~al.}(2014)\citenamefont
  {Helgaker}, \citenamefont {Jorgensen},\ and\ \citenamefont
  {Olsen}}]{helgaker2014molecular}%
  \BibitemOpen
  \bibfield  {author} {\bibinfo {author} {\bibfnamefont {T.}~\bibnamefont
  {Helgaker}}, \bibinfo {author} {\bibfnamefont {P.}~\bibnamefont {Jorgensen}},
  \ and\ \bibinfo {author} {\bibfnamefont {J.}~\bibnamefont {Olsen}},\
  }\href@noop {} {\emph {\bibinfo {title} {Molecular electronic-structure
  theory}}}\ (\bibinfo  {publisher} {John Wiley \& Sons},\ \bibinfo {year}
  {2014})\BibitemShut {NoStop}%
\bibitem [{\citenamefont {McMurchie}\ and\ \citenamefont
  {Davidson}(1978)}]{mcmurchie1978one}%
  \BibitemOpen
  \bibfield  {author} {\bibinfo {author} {\bibfnamefont {L.~E.}\ \bibnamefont
  {McMurchie}}\ and\ \bibinfo {author} {\bibfnamefont {E.~R.}\ \bibnamefont
  {Davidson}},\ }\href@noop {} {\bibfield  {journal} {\bibinfo  {journal}
  {Journal of Computational Physics}\ }\textbf {\bibinfo {volume} {26}},\
  \bibinfo {pages} {218} (\bibinfo {year} {1978})}\BibitemShut {NoStop}%
\bibitem [{\citenamefont {Obara}\ and\ \citenamefont
  {Saika}(1986)}]{obara1986efficient}%
  \BibitemOpen
  \bibfield  {author} {\bibinfo {author} {\bibfnamefont {S.}~\bibnamefont
  {Obara}}\ and\ \bibinfo {author} {\bibfnamefont {A.}~\bibnamefont {Saika}},\
  }\href@noop {} {\bibfield  {journal} {\bibinfo  {journal} {The Journal of
  chemical physics}\ }\textbf {\bibinfo {volume} {84}},\ \bibinfo {pages}
  {3963} (\bibinfo {year} {1986})}\BibitemShut {NoStop}%
\bibitem [{\citenamefont {Friesner}(1985)}]{friesner1985solution}%
  \BibitemOpen
  \bibfield  {author} {\bibinfo {author} {\bibfnamefont {R.~A.}\ \bibnamefont
  {Friesner}},\ }\href@noop {} {\bibfield  {journal} {\bibinfo  {journal}
  {Chemical physics letters}\ }\textbf {\bibinfo {volume} {116}},\ \bibinfo
  {pages} {39} (\bibinfo {year} {1985})}\BibitemShut {NoStop}%
\bibitem [{\citenamefont {Dunlap}\ \emph {et~al.}(1979)\citenamefont {Dunlap},
  \citenamefont {Connolly},\ and\ \citenamefont {Sabin}}]{dunlap1979some}%
  \BibitemOpen
  \bibfield  {author} {\bibinfo {author} {\bibfnamefont {B.~I.}\ \bibnamefont
  {Dunlap}}, \bibinfo {author} {\bibfnamefont {J.}~\bibnamefont {Connolly}}, \
  and\ \bibinfo {author} {\bibfnamefont {J.}~\bibnamefont {Sabin}},\
  }\href@noop {} {\bibfield  {journal} {\bibinfo  {journal} {The Journal of
  Chemical Physics}\ }\textbf {\bibinfo {volume} {71}},\ \bibinfo {pages}
  {3396} (\bibinfo {year} {1979})}\BibitemShut {NoStop}%
\bibitem [{\citenamefont {Dunlap}(1983)}]{dunlap1983fitting}%
  \BibitemOpen
  \bibfield  {author} {\bibinfo {author} {\bibfnamefont {B.}~\bibnamefont
  {Dunlap}},\ }\href@noop {} {\bibfield  {journal} {\bibinfo  {journal} {The
  Journal of Chemical Physics}\ }\textbf {\bibinfo {volume} {78}},\ \bibinfo
  {pages} {3140} (\bibinfo {year} {1983})}\BibitemShut {NoStop}%
\bibitem [{\citenamefont {Mintmire}\ and\ \citenamefont
  {Dunlap}(1982)}]{mintmire1982fitting}%
  \BibitemOpen
  \bibfield  {author} {\bibinfo {author} {\bibfnamefont {J.}~\bibnamefont
  {Mintmire}}\ and\ \bibinfo {author} {\bibfnamefont {B.}~\bibnamefont
  {Dunlap}},\ }\href@noop {} {\bibfield  {journal} {\bibinfo  {journal}
  {Physical Review A}\ }\textbf {\bibinfo {volume} {25}},\ \bibinfo {pages}
  {88} (\bibinfo {year} {1982})}\BibitemShut {NoStop}%
\bibitem [{\citenamefont {Dunlap}(2000)}]{dunlap2000robust}%
  \BibitemOpen
  \bibfield  {author} {\bibinfo {author} {\bibfnamefont {B.}~\bibnamefont
  {Dunlap}},\ }\href@noop {} {\bibfield  {journal} {\bibinfo  {journal}
  {Journal of Molecular Structure: THEOCHEM}\ }\textbf {\bibinfo {volume}
  {529}},\ \bibinfo {pages} {37} (\bibinfo {year} {2000})}\BibitemShut
  {NoStop}%
\bibitem [{\citenamefont {Termath}\ and\ \citenamefont
  {Handy}(1994)}]{Termath1994b}%
  \BibitemOpen
  \bibfield  {author} {\bibinfo {author} {\bibfnamefont {V.}~\bibnamefont
  {Termath}}\ and\ \bibinfo {author} {\bibfnamefont {N.~C.}\ \bibnamefont
  {Handy}},\ }\href {\doibase 10.1016/0009-2614(94)01160-5} {\bibfield
  {journal} {\bibinfo  {journal} {Chemical Physics Letters}\ }\textbf {\bibinfo
  {volume} {230}},\ \bibinfo {pages} {17} (\bibinfo {year} {1994})}\BibitemShut
  {NoStop}%
\bibitem [{\citenamefont {CP975667RF32}\ and\ \citenamefont
  {Rokhlin}(1987)}]{cp975667rf321987fast}%
  \BibitemOpen
  \bibfield  {author} {\bibinfo {author} {\bibfnamefont {L.~G.}\ \bibnamefont
  {CP975667RF32}}\ and\ \bibinfo {author} {\bibfnamefont {V.}~\bibnamefont
  {Rokhlin}},\ }\href@noop {} {\bibfield  {journal} {\bibinfo  {journal} {J.
  Comput. Phys}\ }\textbf {\bibinfo {volume} {73}},\ \bibinfo {pages} {325}
  (\bibinfo {year} {1987})}\BibitemShut {NoStop}%
\bibitem [{\citenamefont {Greengard}(1988)}]{greengard1988rapid}%
  \BibitemOpen
  \bibfield  {author} {\bibinfo {author} {\bibfnamefont {L.}~\bibnamefont
  {Greengard}},\ }\href@noop {} {\emph {\bibinfo {title} {The rapid evaluation
  of potential fields in particle systems}}}\ (\bibinfo  {publisher} {MIT
  press},\ \bibinfo {year} {1988})\BibitemShut {NoStop}%
\bibitem [{\citenamefont {Appel}(1985)}]{appel1985efficient}%
  \BibitemOpen
  \bibfield  {author} {\bibinfo {author} {\bibfnamefont {A.~W.}\ \bibnamefont
  {Appel}},\ }\href@noop {} {\bibfield  {journal} {\bibinfo  {journal} {SIAM
  Journal on Scientific and Statistical Computing}\ }\textbf {\bibinfo {volume}
  {6}},\ \bibinfo {pages} {85} (\bibinfo {year} {1985})}\BibitemShut {NoStop}%
\bibitem [{\citenamefont {Barnes}\ and\ \citenamefont
  {Hut}(1986)}]{barnes1986hierarchical}%
  \BibitemOpen
  \bibfield  {author} {\bibinfo {author} {\bibfnamefont {J.}~\bibnamefont
  {Barnes}}\ and\ \bibinfo {author} {\bibfnamefont {P.}~\bibnamefont {Hut}},\
  }\href@noop {} {\bibfield  {journal} {\bibinfo  {journal} {nature}\ }\textbf
  {\bibinfo {volume} {324}},\ \bibinfo {pages} {446} (\bibinfo {year}
  {1986})}\BibitemShut {NoStop}%
\bibitem [{\citenamefont {Strain}\ \emph {et~al.}(1996)\citenamefont {Strain},
  \citenamefont {Scuseria},\ and\ \citenamefont
  {Frisch}}]{strain1996achieving}%
  \BibitemOpen
  \bibfield  {author} {\bibinfo {author} {\bibfnamefont {M.~C.}\ \bibnamefont
  {Strain}}, \bibinfo {author} {\bibfnamefont {G.~E.}\ \bibnamefont
  {Scuseria}}, \ and\ \bibinfo {author} {\bibfnamefont {M.~J.}\ \bibnamefont
  {Frisch}},\ }\href@noop {} {\bibfield  {journal} {\bibinfo  {journal}
  {Science}\ }\textbf {\bibinfo {volume} {271}},\ \bibinfo {pages} {51}
  (\bibinfo {year} {1996})}\BibitemShut {NoStop}%
\bibitem [{\citenamefont {Schmidt}\ and\ \citenamefont
  {Lee}(1991)}]{schmidt1991implementing}%
  \BibitemOpen
  \bibfield  {author} {\bibinfo {author} {\bibfnamefont {K.~E.}\ \bibnamefont
  {Schmidt}}\ and\ \bibinfo {author} {\bibfnamefont {M.~A.}\ \bibnamefont
  {Lee}},\ }\href@noop {} {\bibfield  {journal} {\bibinfo  {journal} {Journal
  of Statistical Physics}\ }\textbf {\bibinfo {volume} {63}},\ \bibinfo {pages}
  {1223} (\bibinfo {year} {1991})}\BibitemShut {NoStop}%
\bibitem [{\citenamefont {Becke}(1988)}]{Becke1988e}%
  \BibitemOpen
  \bibfield  {author} {\bibinfo {author} {\bibfnamefont {A.~D.}\ \bibnamefont
  {Becke}},\ }\href {\doibase 10.1063/1.454033} {\bibfield  {journal} {\bibinfo
   {journal} {The Journal of Chemical Physics}\ }\textbf {\bibinfo {volume}
  {88}},\ \bibinfo {pages} {2547} (\bibinfo {year} {1988})}\BibitemShut
  {NoStop}%
\bibitem [{\citenamefont {Delley}(1990)}]{Delley1990b}%
  \BibitemOpen
  \bibfield  {author} {\bibinfo {author} {\bibfnamefont {B.}~\bibnamefont
  {Delley}},\ }\href {\doibase 10.1063/1.458452} {\bibfield  {journal}
  {\bibinfo  {journal} {The Journal of Chemical Physics}\ }\textbf {\bibinfo
  {volume} {92}},\ \bibinfo {pages} {508} (\bibinfo {year} {1990})}\BibitemShut
  {NoStop}%
\bibitem [{\citenamefont {Valeev}(2021)}]{Libint2}%
  \BibitemOpen
  \bibfield  {author} {\bibinfo {author} {\bibfnamefont {E.~F.}\ \bibnamefont
  {Valeev}},\ }\href@noop {} {\enquote {\bibinfo {title} {Libint: A library for
  the evaluation of molecular integrals of many-body operators over gaussian
  functions},}\ }\bibinfo {howpublished} {http://libint.valeyev.net/} (\bibinfo
  {year} {2021}),\ \bibinfo {note} {version 2.7.1}\BibitemShut {NoStop}%
\bibitem [{\citenamefont {Sun}(2015)}]{libcint2015}%
  \BibitemOpen
  \bibfield  {author} {\bibinfo {author} {\bibfnamefont {Q.}~\bibnamefont
  {Sun}},\ }\href {\doibase 10.1002/jcc.23981} {\bibfield  {journal} {\bibinfo
  {journal} {Journal of Computational Chemistry}\ }\textbf {\bibinfo {volume}
  {36}},\ \bibinfo {pages} {1664} (\bibinfo {year} {2015})}\BibitemShut
  {NoStop}%
\bibitem [{\citenamefont {Jackson}(1975)}]{Jackson:100964}%
  \BibitemOpen
  \bibfield  {author} {\bibinfo {author} {\bibfnamefont {J.~D.}\ \bibnamefont
  {Jackson}},\ }\href {https://cds.cern.ch/record/100964} {\emph {\bibinfo
  {title} {{Classical electrodynamics; 2nd ed.}}}}\ (\bibinfo  {publisher}
  {Wiley},\ \bibinfo {address} {New York, NY},\ \bibinfo {year}
  {1975})\BibitemShut {NoStop}%
\bibitem [{\citenamefont {Becke}\ and\ \citenamefont
  {Dickson}(1988)}]{Becke1988d}%
  \BibitemOpen
  \bibfield  {author} {\bibinfo {author} {\bibfnamefont {A.~D.}\ \bibnamefont
  {Becke}}\ and\ \bibinfo {author} {\bibfnamefont {R.~M.}\ \bibnamefont
  {Dickson}},\ }\href {\doibase 10.1063/1.455005} {\bibfield  {journal}
  {\bibinfo  {journal} {The Journal of Chemical Physics}\ }\textbf {\bibinfo
  {volume} {89}},\ \bibinfo {pages} {2993} (\bibinfo {year}
  {1988})}\BibitemShut {NoStop}%
\bibitem [{\citenamefont {Posada}\ \emph {et~al.}(2018)\citenamefont {Posada},
  \citenamefont {Moncada},\ and\ \citenamefont {Reyes}}]{Posada2018b}%
  \BibitemOpen
  \bibfield  {author} {\bibinfo {author} {\bibfnamefont {E.}~\bibnamefont
  {Posada}}, \bibinfo {author} {\bibfnamefont {F.}~\bibnamefont {Moncada}}, \
  and\ \bibinfo {author} {\bibfnamefont {A.}~\bibnamefont {Reyes}},\ }\href
  {\doibase 10.1063/1.5012521} {\bibfield  {journal} {\bibinfo  {journal}
  {Journal of Chemical Physics}\ }\textbf {\bibinfo {volume} {148}} (\bibinfo
  {year} {2018}),\ 10.1063/1.5012521}\BibitemShut {NoStop}%
\bibitem [{\citenamefont {Shiozaki}\ and\ \citenamefont
  {Hirata}(2007)}]{Shiozaki2007b}%
  \BibitemOpen
  \bibfield  {author} {\bibinfo {author} {\bibfnamefont {T.}~\bibnamefont
  {Shiozaki}}\ and\ \bibinfo {author} {\bibfnamefont {S.}~\bibnamefont
  {Hirata}},\ }\href {\doibase 10.1103/PhysRevA.76.040503} {\bibfield
  {journal} {\bibinfo  {journal} {Physical Review A - Atomic, Molecular, and
  Optical Physics}\ }\textbf {\bibinfo {volume} {76}},\ \bibinfo {pages} {1}
  (\bibinfo {year} {2007})}\BibitemShut {NoStop}%
\bibitem [{\citenamefont {Becke}(1989)}]{becke1989basis}%
  \BibitemOpen
  \bibfield  {author} {\bibinfo {author} {\bibfnamefont {A.~D.}\ \bibnamefont
  {Becke}},\ }\href@noop {} {\bibfield  {journal} {\bibinfo  {journal}
  {International Journal of Quantum Chemistry}\ }\textbf {\bibinfo {volume}
  {36}},\ \bibinfo {pages} {599} (\bibinfo {year} {1989})}\BibitemShut
  {NoStop}%
\bibitem [{\citenamefont {Becke}(2016)}]{becke2016vertical}%
  \BibitemOpen
  \bibfield  {author} {\bibinfo {author} {\bibfnamefont {A.~D.}\ \bibnamefont
  {Becke}},\ }\href@noop {} {\bibfield  {journal} {\bibinfo  {journal} {The
  Journal of Chemical Physics}\ }\textbf {\bibinfo {volume} {145}},\ \bibinfo
  {pages} {194107} (\bibinfo {year} {2016})}\BibitemShut {NoStop}%
\bibitem [{\citenamefont {Becke}(2018)}]{becke2018communication}%
  \BibitemOpen
  \bibfield  {author} {\bibinfo {author} {\bibfnamefont {A.~D.}\ \bibnamefont
  {Becke}},\ }\href@noop {} {\bibfield  {journal} {\bibinfo  {journal} {The
  Journal of Chemical Physics}\ }\textbf {\bibinfo {volume} {149}},\ \bibinfo
  {pages} {081102} (\bibinfo {year} {2018})}\BibitemShut {NoStop}%
\bibitem [{\citenamefont {Becke}(2019)}]{becke2019dependence}%
  \BibitemOpen
  \bibfield  {author} {\bibinfo {author} {\bibfnamefont {A.~D.}\ \bibnamefont
  {Becke}},\ }\href@noop {} {\bibfield  {journal} {\bibinfo  {journal} {The
  Journal of Chemical Physics}\ }\textbf {\bibinfo {volume} {150}},\ \bibinfo
  {pages} {241101} (\bibinfo {year} {2019})}\BibitemShut {NoStop}%
\bibitem [{\citenamefont {Franchini}\ \emph {et~al.}(2014)\citenamefont
  {Franchini}, \citenamefont {Philipsen}, \citenamefont {{Van Lenthe}},\ and\
  \citenamefont {Visscher}}]{Franchini2014b}%
  \BibitemOpen
  \bibfield  {author} {\bibinfo {author} {\bibfnamefont {M.}~\bibnamefont
  {Franchini}}, \bibinfo {author} {\bibfnamefont {P.~H.~T.}\ \bibnamefont
  {Philipsen}}, \bibinfo {author} {\bibfnamefont {E.}~\bibnamefont {{Van
  Lenthe}}}, \ and\ \bibinfo {author} {\bibfnamefont {L.}~\bibnamefont
  {Visscher}},\ }\href {\doibase 10.1021/ct500172n} {\bibfield  {journal}
  {\bibinfo  {journal} {Journal of Chemical Theory and Computation}\ }\textbf
  {\bibinfo {volume} {10}},\ \bibinfo {pages} {1994} (\bibinfo {year}
  {2014})}\BibitemShut {NoStop}%
\bibitem [{\citenamefont {Blum}\ \emph {et~al.}(2009)\citenamefont {Blum},
  \citenamefont {Gehrke}, \citenamefont {Hanke}, \citenamefont {Havu},
  \citenamefont {Havu}, \citenamefont {Ren}, \citenamefont {Reuter},\ and\
  \citenamefont {Scheffler}}]{Blum2009b}%
  \BibitemOpen
  \bibfield  {author} {\bibinfo {author} {\bibfnamefont {V.}~\bibnamefont
  {Blum}}, \bibinfo {author} {\bibfnamefont {R.}~\bibnamefont {Gehrke}},
  \bibinfo {author} {\bibfnamefont {F.}~\bibnamefont {Hanke}}, \bibinfo
  {author} {\bibfnamefont {P.}~\bibnamefont {Havu}}, \bibinfo {author}
  {\bibfnamefont {V.}~\bibnamefont {Havu}}, \bibinfo {author} {\bibfnamefont
  {X.}~\bibnamefont {Ren}}, \bibinfo {author} {\bibfnamefont {K.}~\bibnamefont
  {Reuter}}, \ and\ \bibinfo {author} {\bibfnamefont {M.}~\bibnamefont
  {Scheffler}},\ }\href {\doibase 10.1016/j.cpc.2009.06.022} {\bibfield
  {journal} {\bibinfo  {journal} {Computer Physics Communications}\ }\textbf
  {\bibinfo {volume} {180}},\ \bibinfo {pages} {2175} (\bibinfo {year}
  {2009})}\BibitemShut {NoStop}%
\bibitem [{\citenamefont {Te~Velde}\ \emph {et~al.}(2001)\citenamefont
  {Te~Velde}, \citenamefont {Bickelhaupt}, \citenamefont {Baerends},
  \citenamefont {Fonseca~Guerra}, \citenamefont {van Gisbergen}, \citenamefont
  {Snijders},\ and\ \citenamefont {Ziegler}}]{te2001chemistry}%
  \BibitemOpen
  \bibfield  {author} {\bibinfo {author} {\bibfnamefont {G.~t.}\ \bibnamefont
  {Te~Velde}}, \bibinfo {author} {\bibfnamefont {F.~M.}\ \bibnamefont
  {Bickelhaupt}}, \bibinfo {author} {\bibfnamefont {E.~J.}\ \bibnamefont
  {Baerends}}, \bibinfo {author} {\bibfnamefont {C.}~\bibnamefont
  {Fonseca~Guerra}}, \bibinfo {author} {\bibfnamefont {S.~J.}\ \bibnamefont
  {van Gisbergen}}, \bibinfo {author} {\bibfnamefont {J.~G.}\ \bibnamefont
  {Snijders}}, \ and\ \bibinfo {author} {\bibfnamefont {T.}~\bibnamefont
  {Ziegler}},\ }\href@noop {} {\bibfield  {journal} {\bibinfo  {journal}
  {Journal of Computational Chemistry}\ }\textbf {\bibinfo {volume} {22}},\
  \bibinfo {pages} {931} (\bibinfo {year} {2001})}\BibitemShut {NoStop}%
\bibitem [{\citenamefont {Franchini}\ \emph {et~al.}(2013)\citenamefont
  {Franchini}, \citenamefont {Philipsen},\ and\ \citenamefont
  {Visscher}}]{Franchini2013b}%
  \BibitemOpen
  \bibfield  {author} {\bibinfo {author} {\bibfnamefont {M.}~\bibnamefont
  {Franchini}}, \bibinfo {author} {\bibfnamefont {P.~H.~T.}\ \bibnamefont
  {Philipsen}}, \ and\ \bibinfo {author} {\bibfnamefont {L.}~\bibnamefont
  {Visscher}},\ }\href {\doibase 10.1002/jcc.23323} {\bibfield  {journal}
  {\bibinfo  {journal} {Journal of Computational Chemistry}\ }\textbf {\bibinfo
  {volume} {34}},\ \bibinfo {pages} {1819} (\bibinfo {year}
  {2013})}\BibitemShut {NoStop}%
\bibitem [{\citenamefont {Te~Velde}\ and\ \citenamefont
  {Baerends}(1991)}]{te1991precise}%
  \BibitemOpen
  \bibfield  {author} {\bibinfo {author} {\bibfnamefont {G.}~\bibnamefont
  {Te~Velde}}\ and\ \bibinfo {author} {\bibfnamefont {E.}~\bibnamefont
  {Baerends}},\ }\href@noop {} {\bibfield  {journal} {\bibinfo  {journal}
  {Physical Review B}\ }\textbf {\bibinfo {volume} {44}},\ \bibinfo {pages}
  {7888} (\bibinfo {year} {1991})}\BibitemShut {NoStop}%
\bibitem [{\citenamefont {Zope}\ \emph {et~al.}(2019)\citenamefont {Zope},
  \citenamefont {Yamamoto}, \citenamefont {Diaz}, \citenamefont {Baruah},
  \citenamefont {Peralta}, \citenamefont {Jackson}, \citenamefont {Santra},\
  and\ \citenamefont {Perdew}}]{zope2019lsic}%
  \BibitemOpen
  \bibfield  {author} {\bibinfo {author} {\bibfnamefont {R.~R.}\ \bibnamefont
  {Zope}}, \bibinfo {author} {\bibfnamefont {Y.}~\bibnamefont {Yamamoto}},
  \bibinfo {author} {\bibfnamefont {C.~M.}\ \bibnamefont {Diaz}}, \bibinfo
  {author} {\bibfnamefont {T.}~\bibnamefont {Baruah}}, \bibinfo {author}
  {\bibfnamefont {J.~E.}\ \bibnamefont {Peralta}}, \bibinfo {author}
  {\bibfnamefont {K.~A.}\ \bibnamefont {Jackson}}, \bibinfo {author}
  {\bibfnamefont {B.}~\bibnamefont {Santra}}, \ and\ \bibinfo {author}
  {\bibfnamefont {J.~P.}\ \bibnamefont {Perdew}},\ }\href@noop {} {\bibfield
  {journal} {\bibinfo  {journal} {The Journal of Chemical Physics}\ }\textbf
  {\bibinfo {volume} {151}},\ \bibinfo {pages} {214108} (\bibinfo {year}
  {2019})}\BibitemShut {NoStop}%
\bibitem [{\citenamefont {Lebedev}(1976)}]{lebedev1976quadratures}%
  \BibitemOpen
  \bibfield  {author} {\bibinfo {author} {\bibfnamefont {V.~I.}\ \bibnamefont
  {Lebedev}},\ }\href@noop {} {\bibfield  {journal} {\bibinfo  {journal} {USSR
  Computational Mathematics and Mathematical Physics}\ }\textbf {\bibinfo
  {volume} {16}},\ \bibinfo {pages} {10} (\bibinfo {year} {1976})}\BibitemShut
  {NoStop}%
\bibitem [{\citenamefont {El-Sherbiny}\ and\ \citenamefont
  {Poirier}(2004)}]{El-Sherbiny2004b}%
  \BibitemOpen
  \bibfield  {author} {\bibinfo {author} {\bibfnamefont {A.}~\bibnamefont
  {El-Sherbiny}}\ and\ \bibinfo {author} {\bibfnamefont {R.~A.}\ \bibnamefont
  {Poirier}},\ }\href {\doibase 10.1002/jcc.20063} {\bibfield  {journal}
  {\bibinfo  {journal} {Journal of Computational Chemistry}\ }\textbf {\bibinfo
  {volume} {25}},\ \bibinfo {pages} {1378} (\bibinfo {year}
  {2004})}\BibitemShut {NoStop}%
\bibitem [{\citenamefont {Gill}\ \emph {et~al.}(1993)\citenamefont {Gill},
  \citenamefont {Johnson},\ and\ \citenamefont {Pople}}]{Gill1993b}%
  \BibitemOpen
  \bibfield  {author} {\bibinfo {author} {\bibfnamefont {P.~M.}\ \bibnamefont
  {Gill}}, \bibinfo {author} {\bibfnamefont {B.~G.}\ \bibnamefont {Johnson}}, \
  and\ \bibinfo {author} {\bibfnamefont {J.~A.}\ \bibnamefont {Pople}},\ }\href
  {\doibase 10.1016/0009-2614(93)80125-9} {\bibfield  {journal} {\bibinfo
  {journal} {Chemical Physics Letters}\ }\textbf {\bibinfo {volume} {209}},\
  \bibinfo {pages} {506} (\bibinfo {year} {1993})}\BibitemShut {NoStop}%
\bibitem [{\citenamefont {Gill}\ and\ \citenamefont {Chien}(2003)}]{Gill2003b}%
  \BibitemOpen
  \bibfield  {author} {\bibinfo {author} {\bibfnamefont {P.~M.}\ \bibnamefont
  {Gill}}\ and\ \bibinfo {author} {\bibfnamefont {S.~H.}\ \bibnamefont
  {Chien}},\ }\href {\doibase 10.1002/jcc.10211} {\bibfield  {journal}
  {\bibinfo  {journal} {Journal of Computational Chemistry}\ }\textbf {\bibinfo
  {volume} {24}},\ \bibinfo {pages} {732} (\bibinfo {year} {2003})}\BibitemShut
  {NoStop}%
\bibitem [{\citenamefont {Kakhiani}\ \emph {et~al.}(2009)\citenamefont
  {Kakhiani}, \citenamefont {Tsereteli},\ and\ \citenamefont
  {Tsereteli}}]{Kakhiani2009b}%
  \BibitemOpen
  \bibfield  {author} {\bibinfo {author} {\bibfnamefont {K.}~\bibnamefont
  {Kakhiani}}, \bibinfo {author} {\bibfnamefont {K.}~\bibnamefont {Tsereteli}},
  \ and\ \bibinfo {author} {\bibfnamefont {P.}~\bibnamefont {Tsereteli}},\
  }\href {\doibase 10.1016/j.cpc.2008.10.004} {\bibfield  {journal} {\bibinfo
  {journal} {Computer Physics Communications}\ }\textbf {\bibinfo {volume}
  {180}},\ \bibinfo {pages} {256} (\bibinfo {year} {2009})}\BibitemShut
  {NoStop}%
\bibitem [{\citenamefont {Krack}\ and\ \citenamefont
  {K{\"{o}}ster}(1998)}]{Krack1998b}%
  \BibitemOpen
  \bibfield  {author} {\bibinfo {author} {\bibfnamefont {M.}~\bibnamefont
  {Krack}}\ and\ \bibinfo {author} {\bibfnamefont {A.~M.}\ \bibnamefont
  {K{\"{o}}ster}},\ }\href {\doibase 10.1063/1.475719} {\bibfield  {journal}
  {\bibinfo  {journal} {Journal of Chemical Physics}\ }\textbf {\bibinfo
  {volume} {108}},\ \bibinfo {pages} {3226} (\bibinfo {year}
  {1998})}\BibitemShut {NoStop}%
\bibitem [{\citenamefont {Lindh}\ \emph {et~al.}(2001)\citenamefont {Lindh},
  \citenamefont {Malmqvist},\ and\ \citenamefont {Gagliardi}}]{Lindh2001b}%
  \BibitemOpen
  \bibfield  {author} {\bibinfo {author} {\bibfnamefont {R.}~\bibnamefont
  {Lindh}}, \bibinfo {author} {\bibfnamefont {P.~{\AA}.}\ \bibnamefont
  {Malmqvist}}, \ and\ \bibinfo {author} {\bibfnamefont {L.}~\bibnamefont
  {Gagliardi}},\ }\href {\doibase 10.1007/s002140100263} {\bibfield  {journal}
  {\bibinfo  {journal} {Theoretical Chemistry Accounts}\ }\textbf {\bibinfo
  {volume} {106}},\ \bibinfo {pages} {178} (\bibinfo {year}
  {2001})}\BibitemShut {NoStop}%
\bibitem [{\citenamefont {Mura}\ and\ \citenamefont
  {Knowles}(1996)}]{Mura1996b}%
  \BibitemOpen
  \bibfield  {author} {\bibinfo {author} {\bibfnamefont {M.~E.}\ \bibnamefont
  {Mura}}\ and\ \bibinfo {author} {\bibfnamefont {P.~J.}\ \bibnamefont
  {Knowles}},\ }\href {\doibase 10.1063/1.471749} {\bibfield  {journal}
  {\bibinfo  {journal} {Journal of Chemical Physics}\ }\textbf {\bibinfo
  {volume} {104}},\ \bibinfo {pages} {9848} (\bibinfo {year}
  {1996})}\BibitemShut {NoStop}%
\bibitem [{\citenamefont {Murray}\ \emph {et~al.}(1993)\citenamefont {Murray},
  \citenamefont {Handy},\ and\ \citenamefont {Laming}}]{Murray1993b}%
  \BibitemOpen
  \bibfield  {author} {\bibinfo {author} {\bibfnamefont {C.~W.}\ \bibnamefont
  {Murray}}, \bibinfo {author} {\bibfnamefont {N.~C.}\ \bibnamefont {Handy}}, \
  and\ \bibinfo {author} {\bibfnamefont {G.~J.}\ \bibnamefont {Laming}},\
  }\href {\doibase 10.1080/00268979300100651} {\bibfield  {journal} {\bibinfo
  {journal} {Molecular Physics}\ }\textbf {\bibinfo {volume} {78}},\ \bibinfo
  {pages} {997} (\bibinfo {year} {1993})}\BibitemShut {NoStop}%
\bibitem [{\citenamefont {Treutler}\ and\ \citenamefont
  {Ahlrichs}(1995)}]{Treutler1995b}%
  \BibitemOpen
  \bibfield  {author} {\bibinfo {author} {\bibfnamefont {O.}~\bibnamefont
  {Treutler}}\ and\ \bibinfo {author} {\bibfnamefont {R.}~\bibnamefont
  {Ahlrichs}},\ }\href {\doibase 10.1063/1.469408} {\bibfield  {journal}
  {\bibinfo  {journal} {The Journal of Chemical Physics}\ }\textbf {\bibinfo
  {volume} {102}},\ \bibinfo {pages} {346} (\bibinfo {year}
  {1995})}\BibitemShut {NoStop}%
\bibitem [{\citenamefont {Stratmann}\ \emph {et~al.}(1996)\citenamefont
  {Stratmann}, \citenamefont {Scuseria},\ and\ \citenamefont
  {Frisch}}]{Stratmann1996b}%
  \BibitemOpen
  \bibfield  {author} {\bibinfo {author} {\bibfnamefont {R.~E.}\ \bibnamefont
  {Stratmann}}, \bibinfo {author} {\bibfnamefont {G.~E.}\ \bibnamefont
  {Scuseria}}, \ and\ \bibinfo {author} {\bibfnamefont {M.~J.}\ \bibnamefont
  {Frisch}},\ }\href {\doibase 10.1016/0009-2614(96)00600-8} {\bibfield
  {journal} {\bibinfo  {journal} {Chemical Physics Letters}\ }\textbf {\bibinfo
  {volume} {257}},\ \bibinfo {pages} {213} (\bibinfo {year}
  {1996})}\BibitemShut {NoStop}%
\bibitem [{\citenamefont {Gharibnejad}\ \emph {et~al.}(2021)\citenamefont
  {Gharibnejad}, \citenamefont {Douguet}, \citenamefont {Schneider},
  \citenamefont {Olsen},\ and\ \citenamefont {Argenti}}]{Gharibnejad2021b}%
  \BibitemOpen
  \bibfield  {author} {\bibinfo {author} {\bibfnamefont {H.}~\bibnamefont
  {Gharibnejad}}, \bibinfo {author} {\bibfnamefont {N.}~\bibnamefont
  {Douguet}}, \bibinfo {author} {\bibfnamefont {B.~I.}\ \bibnamefont
  {Schneider}}, \bibinfo {author} {\bibfnamefont {J.}~\bibnamefont {Olsen}}, \
  and\ \bibinfo {author} {\bibfnamefont {L.}~\bibnamefont {Argenti}},\ }\href
  {\doibase 10.1016/j.cpc.2021.107889} {\bibfield  {journal} {\bibinfo
  {journal} {Computer Physics Communications}\ }\textbf {\bibinfo {volume}
  {263}} (\bibinfo {year} {2021}),\ 10.1016/j.cpc.2021.107889},\ \Eprint
  {http://arxiv.org/abs/2101.08678} {arXiv:2101.08678} \BibitemShut {NoStop}%
\bibitem [{\citenamefont {Laqua}\ \emph {et~al.}(2018)\citenamefont {Laqua},
  \citenamefont {Kussmann},\ and\ \citenamefont {Ochsenfeld}}]{Laqua2018b}%
  \BibitemOpen
  \bibfield  {author} {\bibinfo {author} {\bibfnamefont {H.}~\bibnamefont
  {Laqua}}, \bibinfo {author} {\bibfnamefont {J.}~\bibnamefont {Kussmann}}, \
  and\ \bibinfo {author} {\bibfnamefont {C.}~\bibnamefont {Ochsenfeld}},\
  }\href {\doibase 10.1063/1.5049435} {\bibfield  {journal} {\bibinfo
  {journal} {Journal of Chemical Physics}\ }\textbf {\bibinfo {volume} {149}}
  (\bibinfo {year} {2018}),\ 10.1063/1.5049435}\BibitemShut {NoStop}%
\bibitem [{\citenamefont {Arfken}(1985)}]{garfken67:math}%
  \BibitemOpen
  \bibfield  {author} {\bibinfo {author} {\bibfnamefont {G.}~\bibnamefont
  {Arfken}},\ }\href@noop {} {\emph {\bibinfo {title} {Mathematical Methods for
  Physicists}}},\ \bibinfo {edition} {3rd}\ ed.\ (\bibinfo  {publisher}
  {Academic Press, {Inc.}},\ \bibinfo {address} {San Diego},\ \bibinfo {year}
  {1985})\BibitemShut {NoStop}%
\bibitem [{\citenamefont {Porezag}\ and\ \citenamefont
  {Pederson}(1999)}]{Porezag_bsis}%
  \BibitemOpen
  \bibfield  {author} {\bibinfo {author} {\bibfnamefont {D.}~\bibnamefont
  {Porezag}}\ and\ \bibinfo {author} {\bibfnamefont {M.~R.}\ \bibnamefont
  {Pederson}},\ }\href {\doibase 10.1103/PhysRevA.60.2840} {\bibfield
  {journal} {\bibinfo  {journal} {Phys. Rev. A}\ }\textbf {\bibinfo {volume}
  {60}},\ \bibinfo {pages} {2840} (\bibinfo {year} {1999})}\BibitemShut
  {NoStop}%
\bibitem [{\citenamefont {von Winckel}(2022)}]{FDMweight}%
  \BibitemOpen
  \bibfield  {author} {\bibinfo {author} {\bibfnamefont {G.}~\bibnamefont {von
  Winckel}},\ }\href
  {https://www.mathworks.com/matlabcentral/fileexchange/5269-finite-difference-weights}
  {\bibfield  {journal} {\bibinfo  {journal} {Greg von Winckel (2022). Finite
  Difference Weights
  (https://www.mathworks.com/matlabcentral/fileexchange/5269-finite-difference-weights),
  MATLAB Central File Exchange. Retrieved February 2, 2022.}\ } (\bibinfo
  {year} {2022})}\BibitemShut {NoStop}%
\bibitem [{\citenamefont {Press}\ \emph {et~al.}(1992)\citenamefont {Press},
  \citenamefont {Flannery}, \citenamefont {Teukolsky},\ and\ \citenamefont
  {Vetterling}}]{nrf77}%
  \BibitemOpen
  \bibfield  {author} {\bibinfo {author} {\bibfnamefont {W.~H.}\ \bibnamefont
  {Press}}, \bibinfo {author} {\bibfnamefont {B.~P.}\ \bibnamefont {Flannery}},
  \bibinfo {author} {\bibfnamefont {S.~A.}\ \bibnamefont {Teukolsky}}, \ and\
  \bibinfo {author} {\bibfnamefont {W.~T.}\ \bibnamefont {Vetterling}},\ }\href
  {http://www.worldcat.org/isbn/052143064X} {\emph {\bibinfo {title} {Numerical
  Recipes in FORTRAN 77: The Art of Scientific Computing}}},\ \bibinfo
  {edition} {2nd}\ ed.\ (\bibinfo  {publisher} {Cambridge University Press},\
  \bibinfo {year} {1992})\BibitemShut {NoStop}%
\bibitem [{\citenamefont {Perdew}\ and\ \citenamefont
  {Zunger}(1981)}]{perdew1981pzsic}%
  \BibitemOpen
  \bibfield  {author} {\bibinfo {author} {\bibfnamefont {J.~P.}\ \bibnamefont
  {Perdew}}\ and\ \bibinfo {author} {\bibfnamefont {A.}~\bibnamefont
  {Zunger}},\ }\href@noop {} {\bibfield  {journal} {\bibinfo  {journal}
  {Physical Review B}\ }\textbf {\bibinfo {volume} {23}},\ \bibinfo {pages}
  {5048} (\bibinfo {year} {1981})}\BibitemShut {NoStop}%
\bibitem [{\citenamefont {Vydrov}\ \emph {et~al.}(2006)\citenamefont {Vydrov},
  \citenamefont {Scuseria}, \citenamefont {Perdew}, \citenamefont
  {Ruzsinszky},\ and\ \citenamefont {Csonka}}]{vydrov2006osic}%
  \BibitemOpen
  \bibfield  {author} {\bibinfo {author} {\bibfnamefont {O.~A.}\ \bibnamefont
  {Vydrov}}, \bibinfo {author} {\bibfnamefont {G.~E.}\ \bibnamefont
  {Scuseria}}, \bibinfo {author} {\bibfnamefont {J.~P.}\ \bibnamefont
  {Perdew}}, \bibinfo {author} {\bibfnamefont {A.}~\bibnamefont {Ruzsinszky}},
  \ and\ \bibinfo {author} {\bibfnamefont {G.~I.}\ \bibnamefont {Csonka}},\
  }\href@noop {} {\bibfield  {journal} {\bibinfo  {journal} {The Journal of
  chemical physics}\ }\textbf {\bibinfo {volume} {124}},\ \bibinfo {pages}
  {094108} (\bibinfo {year} {2006})}\BibitemShut {NoStop}%
\bibitem [{\citenamefont {Yamamoto}\ \emph {et~al.}(2020)\citenamefont
  {Yamamoto}, \citenamefont {Romero}, \citenamefont {Baruah},\ and\
  \citenamefont {Zope}}]{yamamoto2020sosic}%
  \BibitemOpen
  \bibfield  {author} {\bibinfo {author} {\bibfnamefont {Y.}~\bibnamefont
  {Yamamoto}}, \bibinfo {author} {\bibfnamefont {S.}~\bibnamefont {Romero}},
  \bibinfo {author} {\bibfnamefont {T.}~\bibnamefont {Baruah}}, \ and\ \bibinfo
  {author} {\bibfnamefont {R.~R.}\ \bibnamefont {Zope}},\ }\href@noop {}
  {\bibfield  {journal} {\bibinfo  {journal} {The Journal of Chemical Physics}\
  }\textbf {\bibinfo {volume} {152}},\ \bibinfo {pages} {174112} (\bibinfo
  {year} {2020})}\BibitemShut {NoStop}%
\end{thebibliography}%

\end{document}